\documentclass[hyper,letterpaper]{JHEP3}
\usepackage{amsmath,amssymb,amsfonts,bm}
\usepackage{cite}
\usepackage{graphicx}
\usepackage{multirow}
\usepackage{verbatim}
\usepackage{appendix}
\usepackage{epsfig}

\usepackage[matrix,arrow]{xy}

\usepackage{url}
\usepackage{float}

\newcommand{\bea}{\begin{eqnarray}}
\newcommand{\eea}{\end{eqnarray}}
\newcommand{\be}{\begin{equation}}
\newcommand{\ee}{\end{equation}}

\newcommand{\R}{{\mathbb R}}
\newcommand{\C}{{\mathbb C}}

\newcommand{\nco}{\newcommand}

\nco{\ghat}{\widehat{\mathfrak g}}
\def\gg{{\mathfrak g}}
\nco{\out}{\on{out}}
\nco{\on}{\operatorname}
\nco{\ol}{\overline}
\nco{\CJ}{{\mathcal J}}
\nco{\pa}{\partial}
\nco{\FZ}{{\mathcal Z}}
\nco{\ot}{\otimes}
\nco{\bsl}{\backslash}

\def\a{\alpha}
\def\b{\beta}

\def\la{\lambda}

\def\tilde{\widetilde}
\def\hat{\widehat}

\def\CB{{\mathcal B}}
\def\CC{{\mathcal C}}
\def\CD{{\mathcal D}}
\def\CE{{\mathcal E}}
\def\CF{{\mathcal F}}
\def\CG{{\mathcal G}}

\def\CI{{\mathcal I}}
\def\CJ{{\mathcal J}}
\def\CK{{\mathcal K}}
\def\CL{{\mathcal L}}
\def\CM{{\mathcal M}}
\def\CN{{\mathcal N}}
\def\CO{{\mathcal O}}
\def\CP{{\mathcal P}}

\def\CR{{\mathcal R}}
\def\CS{{\mathcal S}}
\def\CT{{\mathcal T}}
\def\CU{{\mathcal U}}
\def\CV{{\mathcal V}}
\def\CW{{\mathcal W}}

\def\CY{{\mathcal Y}}
\def\CZ{{\mathcal Z}}

\newcommand{\ra}{\to}
\newcommand{\rf}[1]{(\ref{#1})}
\newcommand{\fsl}{{\mathfrak s}{\mathfrak l}}

\newcommand{\al}{\alpha}

\newcommand{\ga}{\gamma}
\newcommand{\Ga}{\Gamma}
\newcommand{\de}{\delta}
\newcommand{\De}{\Delta}
\newcommand{\ep}{\epsilon}

\newcommand{\om}{\omega}

\newcommand{\si}{\sigma}

\newcommand{\vf}{\varphi}

\newcommand{\SH}{{\mathsf H}}

\newcommand{\sk}{{\mathsf k}}
\newcommand{\SL}{{\mathsf L}}

\newcommand{\CWL}{\tilde{\mathcal W}^{\rm\sst L}}

\newcommand{\sa}{{\mathsf a}}

\newcommand{\BB}{{\mathbb B}}

\newcommand{\BD}{{\mathbb D}}
\newcommand{\BP}{{\mathbb P}}
\newcommand{\BC}{{\mathbb C}}
\newcommand{\BZ}{{\mathbb Z}}

\newcommand{\sst}{\scriptscriptstyle}

\renewcommand{\hat}{\widehat}

\title{Surface Operators and Separation of Variables}

\author{Edward Frenkel$^{1}$, Sergei Gukov$^{2,3}$ and J\"org Teschner$^{4}$
\\
$^1$ University of California, Berkeley, CA 94720-3840 USA \\
$^2$ California Institute of Technology, Pasadena, CA 91125, USA \\
$^3$ Simons Center for Geometry and Physics, Stony Brook, NY 11794, USA \\
$^4$ DESY Theory, Notkestr. 85, 22603 Hamburg, Germany}

\abstract{Alday, Gaiotto, and Tachikawa conjectured relations
  between certain 4d $N=2$ supersymmetric field theories and 2d
  Liouville conformal field theory. We study generalizations of these
  relations to 4d theories with surface operators. For one type of
  surface operators the corresponding 2d theory is the WZW model, and
  for another type -- the Liouville theory with insertions of extra
  degenerate fields. We show that these two 4d theories with surface
  operators exhibit an IR duality, which reflects the known relation
  (the so-called separation of variables) between the conformal blocks
  of the WZW model and the Liouville theory. Furthermore, we trace
  this IR duality to a brane creation construction relating systems of
  M5 and M2 branes in M-theory. Finally, we show that this duality may
  be expressed as an explicit relation between the generating
  functions for the changes of variables between natural sets of
  Darboux coordinates on the Hitchin moduli space.
\\
\\
\\
\\
\\
\\
{\tt CALT 2015-032}}

\begin{document}


\section{Introduction}

One of the most interesting phenomena in supersymmetric gauge dynamics
is the appearance of infrared (IR) duality: theories different in the
ultraviolet (UV) regime may well flow to the same IR fixed point.  A
prominent example is the Seiberg duality in four-dimensional $\CN=1$
super-QCD \cite{Seiberg:1994pq}. Similar dualities exist in three
dimensions \cite{Aharony:1997gp,Giveon:2008zn} and in two dimensions
\cite{Hori:2006dk}.  Moreover, it is known that certain
two-dimensional dualities naturally arise on the two-dimensional
world-sheets of surface operators in four-dimensional $\CN=2$ gauge
theories \cite{Gadde:2013dda,Gomis:2014eya}.
In the present paper, we propose a new IR duality between 4d $\CN=2$
supersymmetric theories with two types of surface operators that we call
``codimension-2'' and ``codimension-4'' for reasons that will become
clear momentarily.

In general, in four dimensional gauge theory (with any amount of
supersymmetry) we have two ways of constructing non-local operators
supported on a surface $D \subset M_4$ \cite{Gukov:2006jk}:

\begin{itemize}

\item {\bf 2d-4d system:} One can couple 4d gauge theory on $M_4$ to
  an auxiliary 2d theory on $D$ in such a way that the gauge group $G$
  of the 4d theory is a subgroup of the global flavor symmetry of the
  2d theory.  In particular, the auxiliary 2d theory must have global
  symmetry $G$.

\item {\bf singularity:} One replaces the four-dimensional space-time
  $M_4$ with the complement $M_4 \setminus D$ so that gauge fields
  (and, possibly, other fields) have a prescribed singular behavior
  along $D$. Thus, instead of introducing new degrees of freedom, one
  modifies the existing degrees of freedom.

\end{itemize}

\noindent

Note that both of these methods may also be used to construct other
non-local operators, such as line operators (for example, Wilson
operators and 't Hooft operators, respectively). In the case of
surface operators, the first of these two methods can be further
subdivided into linear and non-linear sigma-model descriptions of 2d
degrees of freedom on $D$. However, this distinction will not be
important in this paper.

What will be important to us, however, is that sometimes these two
constructions may lead to the same result. This happens when
integrating out 2d degrees of freedom in the 2d-4d coupled system
leaves behind a delta-function singularity, supported on $D$ (for the
4d fields).  In particular, this is what one finds in the case of
$\CN=4$ super-Yang-Mills theory. Thus, one obtains an equivalence of
the theories with two types of surface operators, which may also be
derived using brane constructions and T-dualities.  Something similar
may happen in certain gauge theories with less supersymmetry, {\it
  e.g.} free field theories, but in this paper focus on {\em IR
  equivalence} (or IR duality) of 4d $\CN=2$ theories with the two
types of surface operators.

Surface operators in 4d $\CN=2$ theories were first considered in
\cite{Gukov:2007ck} and later incorporated in the framework of the
Alday-Gaiotto-Tachikawa (AGT) correspondence in \cite{AGGTV,AT}
relating a certain class of 4d $\CN=2$ gauge theories (often called
``class $\mathcal{S}$'') and 2d conformal field theories on a Riemann
surface $C_{g,n}$ of genus $g$ with $n$ punctures
\cite{Alday:2009aq}. According to these works, there is a relation
between the instanton partition functions in the 4d theories in the
presence of the two types of surface operators and conformal blocks in
the WZW model for $SL_2$ and the Liouville theory with extra
degenerate fields, respectively. We note that for the surface
operators of the first type this relation was originally proposed by
Braverman \cite{Brav} and further analyzed in
\cite{AT,KPPW,Negut,Nawata}.

Within this framework, the IR duality between the 4d theories with two
types of surface operators is neatly expressed by an integral
transform between the chiral partition functions of the WZW model and
the Liouville theory:
\begin{equation}\label{SOV-1}
\CZ^{\rm\sst WZ}(x,z)\,=\,\int du \;K(x,u)\,\CZ^{\rm\sst L}(u,z)\,,
\end{equation}
This relation, which is of interest in 2d CFT, was established by
Feigin, Frenkel, and Stoyanovsky in 1995 as a generalization of the
Sklyanin separation of variables for the Gaudin model \cite{Skl}
(which corresponds to the limit of the infinite central charge), see
\cite{F:icmp,St}. Hence we call this relation {\em separation of
  variables}. In this paper we present it in a more explicit form (see
\cite{Ribault:2005wp} for another presentation).

One of our goals is thus to show that the relation \eqref{SOV-1}
captures the IR duality of 4d $\CN=2$ gauge theories of class
$\mathcal{S}$ with surface operators. Thus, our work provides a
physical interpretation -- and perhaps a natural home -- for the
separation of variables \eqref{SOV-1} in 4d gauge theory, as well as
the corresponding 6d $(0,2)$ theory on the fivebrane world-volume in
M-theory.

Let's talk about the latter in more detail. In the context of the AGT
correspondence and, more broadly, in 4d $\CN=2$ theories constructed
from M-theory fivebranes wrapped on Riemann surfaces
\cite{Witten:1997sc,Maldacena:2000mw,Gaiotto:2009we,Gaiotto:2009hg}
the two types of surface operators in 4d field theories described
above are usually represented by different types of branes /
supersymmetric defects in the 6d $(0,2)$ theory on the fivebrane
world-volume.  Codimension-4 defects that correspond to the membrane
boundaries naturally lead to the surface operators described as 2d-4d
coupled systems. Codimension-2 defects, on the other hand, may be
thought of as intersections with another group of fivebranes and
therefore they are usually characterized by a singularity for the
gauge fields at $D$ of a specific type (described in Appendix \ref{Nahm}).

Thus, altogether one has at least three different perspectives on the
surface operators in 4d theories corresponding to the codimension-2
and codimension-4 defects in 6d theory (this is the reason why we will
often refer to them as codimension-2 and codimension-4 surface
operators). Namely, the 2d CFT perspective, the 4d gauge theory
perspective, and the 6d fivebrane / M-theory perspective.  Moreover,
the 4d gauge theory perspective
is further subdivided into UV and IR regimes. A simple way to
keep track of these perspectives is to think of a sequence of RG
flows, \be \text{M-theory / 6d} \quad \leadsto \quad \text{4d gauge
  theory UV} \quad \leadsto \quad \text{4d gauge theory IR}
\label{RGflow}
\ee where arrows correspond to integrating out more and more degrees
of freedom.  This relation between different theories is somewhat
analogous to a more familiar relation between a 2d gauged linear
sigma-model, the corresponding non-linear sigma-model, and the
Landau-Ginzburg theory that describes the IR physics of the latter.

It is natural to ask whether one can see any trace of our IR
equivalence in the UV, either in 4d or 6d.  We answer this question in
the affirmative, by showing that the brane configurations in M-theory
that give rise to the codimension-2 and codimension-4 surface
operators are related by a certain non-trivial phase transition, a
variant of the {\it brane creation} effect of Hanany and Witten
\cite{HW} (see Figure \ref{M2creationfig} in Section
\ref{branecon}). We will show that certain quantities protected by
supersymmetry remain invariant under this phase transition, thereby
revealing the 6d / M-theory origin of our IR equivalence. In four
dimensions, the IR duality manifests itself in the most direct way as a
relation between instanton partition functions in the presence of
surface operators and conformal blocks in WZW/Liouville CFTs discussed
above. However, what we actually claim here is that the IR duality
holds for the {\em full physical theories} (and not just for specific
observables); that is to say, the 4d theories with two types of
surface operators become {\em equivalent} in the IR. This has many
useful implications (and applications), far beyond a mere relation
between the instanton partition functions.

In order to show that, we use the fact that the low-energy effective
action in our theories is essentially determined by their respective
effective twisted superpotentials (see Sections \ref{4d descr} and
\ref{gen fn} for more details). Hence we need to compare the twisted
superpotentials arising in our theories, and we compute them
explicitly using the corresponding 2d conformal field theories. The
result is that the two twisted superpotentials, which we denote by
$\tilde{\CW}^{{\text M5}}(a,x,\tau)$ and $\tilde{\CW}^{\text
  M2}(a,u(a,x,\tau),\tau)$, respectively, are related by a field
redefinition\footnote{As usual, it is convenient to think of
  parameters as background fields \cite{Seiberg:1993vc}.}
\begin{equation}
\tilde{\CW}^{{\text M5}}(a,x,\tau) \; = \;
\tilde{\CW}^\text{M2}(a,u(a,x,\tau),\tau) +
\tilde{\CW}^\text{SOV}(x,u(a,x,\tau),\tau)\, .
\label{WWW}
\end{equation}
Here the variables $x$ and $u$ are parameters entering the
UV-definitions of the two types of surface operators. The relation
$u=u(a,x,\tau)$ extremizes the superpotential on the right of \rf{WWW},
reflecting the fact that $u$ becomes a dynamical field in our brane creation transition.

Formula \eqref{WWW} has an elegant interpretation in terms of the
mathematics of the Hitchin integrable system for the group
$SL_2$. Namely, we show that the two effective twisted superpotentials
are the generating functions for changes of variables between natural
sets of Darboux coordinates for the Hitchin moduli space $\CM_{\rm
  \sst H}(C)$ of $SL_2$.

There are in fact three such sets: $(x,p)$, the natural coordinates on
$\CM_{\rm \sst H}(C)$ arising from its realization as a cotangent
bundle; $(a,t)$, the action-angle coordinates making the complete
integrability of $\CM_{\rm \sst H}(C)$ manifest; and $(u,v)$, the so-called
``separated variables'' making the eigenvalue equations of the
quantized Hitchin systems separate. We show that the twisted superpotentials
$\tilde \CW^{{\rm\sst M5}} (a,x,\tau)$ and $\tilde \CW^{\rm\sst M2}
(a,u,\tau)$ are the generating functions for the changes of Darboux
coordinates $(x,p) \leftrightarrow (a,t)$ and $(u,v) \leftrightarrow
(a,t)$, respectively. The generating function of the remaining change
$(x,p) \leftrightarrow (u,v)$ is the function
$\tilde{\CW}^\text{SOV}(x,u,\tau)$ appearing on the RHS of the relation
\eqref{WWW} -- it is the generating function for the separation of
variables in the Hitchin integrable system.

\medskip

$$
\xymatrix{& (x,p) \; \text{coordinates}
  \ar@2{<->}[dd]^{\displaystyle{\tilde{\CW}^\text{SOV}}} \\
(a,t) \; \text{coordinates} \ar@2{<->}[ur]^{\displaystyle{\tilde
    \CW^{{\rm\sst M5}}}} \ar@2{<->}[dr]^{\displaystyle{\tilde
    \CW^{\rm\sst M2}}} \\ & (u,v) \; \text{coordinates}
}
$$

\medskip

Thus, the IR duality between the 4d gauge theories with the two types
of surface operators that we study in this paper becomes directly
reflected in the separation of variables of the Hitchin integrable
system.

To derive the relation \eqref{WWW}, we first express the twisted
superpotentials $\tilde \CW^{{\rm\sst M5}} (a,x,\tau)$ and $\tilde
\CW^{\rm\sst M2} (a,u,\tau)$ as the subleading terms in the expansion
of the logarithms of the instanton partition functions in the limit of
vanishing Omega-deformation \cite{Nekrasov:2002qd}. Assuming that the
instanton partition function in our 4d theories are equal to the chiral
partition functions in the WZW model and the Liouville theory,
respectively \cite{Brav,AT,KPPW,Negut,Nawata}, we express the
subleading terms of the instanton partition functions as the
subleading terms of the chiral partition functions in the
corresponding 2d CFTs. What remains to be done then is to find a
relation between the subleading terms of these two chiral partition
functions (one from the WZW model and one from the Liouville theory
with extra degenerate fields).

This is now a problem in 2d CFT, which is in fact a non-trivial
mathematical problem that is interesting on its own right. In this
paper, by refining earlier observations from \cite{Teschner:2010je},
we compute explicitly the subleading terms of the chiral partition
functions in the WZW model and the Liouville theory (with extra
degenerate fields) and identify them as the generating functions for
the changes of Darboux coordinates mentioned above. In this
way we obtain the desired relation \eqref{WWW}.

The details of these computations are given in the Appendices, which
contain a number of previously unpublished results that could be of
independent interest. In performing these computations, we addressed
various points in the mathematics of the WZW model and its relation to
the Hitchin integrable system that, as far as we know, have not been
discussed in the literature before (for example, questions concerning
chiral partition functions on Riemann surfaces of higher genus).  In
particular, our results make precise the sense in which Liouville
theory and the WZW model both appear as the result of natural
quantizations of the Hitchin integrable systems using two different
sets of Darboux coordinates, as was previously argued in
\cite{Teschner:2010je}.

Once we identify the subleading terms of the chiral partition
functions of the two 2d CFTs with the generating functions, we obtain
the relation \eqref{WWW}. Alternatively, this relation also appears in
the infinite central charge limit from the separation of variables
relation \eqref{SOV-1} between conformal blocks in the WZW and
Liouville CFTs. Therefore, the relation \eqref{SOV-1} may be viewed as
a relation between the instanton partition functions in the 4d
theories with two types of surface operators in non-trivial
Omega-background. This suggests that these two 4d theories remain IR
equivalent even after we turn on the Omega-deformation. However, in
non-zero Omega-background this relation is rather non-trivial, as it
involves not just a change of variables, but also an integral
transform. This relation deserves further study, as does
the question of generalizing our results from the group $SL_2$ to
groups of higher rank.

\bigskip

The paper is organized as follows. In Section 2 we review class $\CS$
supersymmetric gauge theories, AGT correspondence, surface operators,
and the Hitchin system. In Section 3 we discuss the 4d theories with
the surface operators obtained from codimension-2 defects in 6d, the
brane construction, conformal blocks in the corresponding CFT (WZW
model), and the relation to the Hitchin system. In Section 4 we
consider the 4d theories with the surface operators obtained from
codimension-2 defects in 6d and the corresponding CFT (Liouville
theory with degenerate fields). We also discuss general properties of
the 4d theories in the IR regime and the corresponding twisted
superpotentials. Anticipating the IR duality that we establish in this
paper, we start with the brane system introduced in Section 3 (the one
giving rise to the codimension-2 defects) and deform it in such a way
that the end result is a collection of codimension-4 defects. This
allows us to demonstrate that the two types of defects preserve the
same subalgebra of the supersymmetry algebra and to set the stage for
the IR duality. In the second half of Section 4, we bring together the
results of the previous sections to demonstrate the IR duality of two
4d gauge theories with surface operators and the separation of
variables in conformal field theory and Hitchin system.

The necessary mathematical results on surface operators, on chiral
partition functions in the WZW model and the Liouville theory, and on
the separation of variables are presented in the Appendices. There one
can also find detailed computations of the chiral partition
functions of the WZW model and the Liouville theory and their
classical limits (some of which have not appeared in the literature
before, as far as we know).

\subsection{Acknowledgments}
We would like to thank D.~Gaiotto, K.~Maruyoshi, and N.~Nekrasov for
useful discussions and comments. The research of E.F. was supported by
the NSF grants DMS-1160328 and DMS-1201335.  The work of S.G. is
funded in part by the DOE Grant DE-SC0011632 and the Walter Burke
Institute for Theoretical Physics.


\section{Preliminaries}
\label{classS}

In this section we review some background and introduce the notation
that will be used in our paper.  Toward this end, we will recall the
notion of class $\CS$ supersymmetric gauge theories and review very
briefly how the Seiberg-Witten theory of this class is related to the
Hitchin system.

\subsection{Theories of class $\CS$ and AGT correspondence}
\label{classS1}

A lot of progress has been made in the last few years in the study of
$\mathcal{N}=2$ supersymmetric field theories in four dimensions.
Highlights include exact results on the expectation values of
observables like supersymmetric Wilson and 't Hooft loop operators on
the four-sphere $S^4$, see \cite{Pestun:2014mja,Okuda:2014fja} for
reviews, and \cite{Teschner:2014oja} for a general overview containing
further references.

A rich  class of field theories with $\mathcal{N}=2$ supersymmetry,
often denoted as class $\mathcal{S}$, can be obtained by twisted compactification
of the six-dimensional $(2,0)$ theory with Lie algebra $\mathfrak{g}$ \cite{Gaiotto:2009hg}.
Class $\mathcal{S}$ theories of type $\mathfrak{g}=A_1$
have Lagrangian descriptions specified by a pair of pants decompositions of $C$,
which is defined by cutting $C$ along a system $\CC=\{\ga_1,\dots,\ga_h\}$ of simple
closed curves on $C$ \cite{Gaiotto:2009we}. In order to distinguish pants decompositions
that differ by Dehn twists, we will also introduce a trivalent graph $\Gamma$
inside $C$ such that each pair of pants contains exactly one vertex of
$\Gamma$, and each edge $e$ of $\Gamma$ goes through exactly one
cutting curve $\ga_e\in\CC$. The pair $\si=(\CC,\Ga)$ will be called a refined pants decomposition.

Then, to a Riemann surface $C$ of genus $g$ and $n$ punctures one may
associate \cite{Gaiotto:2009we,Gaiotto:2009hg} a four-dimensional gauge theory
$\CG_C$ with $\CN=2$ supersymmetry, gauge group $({\rm SU}(2))^{h}$,
$h:=3g-3+n$ and flavor symmetry $({\rm SU}(2))^n$. The theories in
this class are UV-finite, and therefore they are characterized by a
collection of gauge coupling {\it constants} $g_1,\dots,g_h$.
To the $k$-th boundary there corresponds a flavor group $SU(2)_k$ with
mass parameter $M_k$. The hypermultiplet masses are linear
combinations of the parameters $m_k$, $k=1,\dots,n$ as explained in more
detail in \cite{Gaiotto:2009we,Alday:2009aq}.

The correspondence between the data associated to the surface $C$ and
the gauge theory $\CG_C$ is then summarized in the table below.

We place this in the context of M-theory, following the standard
conventions of brane constructions \cite{Witten:1997sc}. Namely, we
choose $x^6$ and $x^{10}$ as local coordinates on the Riemann surface
$C$ and parametrize the four-dimensional space-time $M_4$ by
$(x^0,x^1,x^2,x^3)$.  This choice of local coordinates can be
conveniently summarized by the diagram:
\begin{center}
\begin{tabular}{l || c|c|c|c|c|c|c|c|c|c|c}
Brane & 0 & 1 & 2 & 3 & 4 & 5 & 6 & 7 & 8 & 9 & 10 \\ \hline\hline
$M5$  & x & x & x & x &   &   & x &   &   &   & x
\end{tabular}
\end{center}
where each ``x'' represents a space-time dimensions spanned by the
five-brane world-volume.

Alday, Gaiotto, and Tachikawa (AGT) observed that the
partition functions of $A_1$ theories on a four-sphere can be expressed in terms of Liouville correlation functions.

\begin{center}
\begin{tabular}{l|l}
Riemann surface $C$ & Gauge theory $\CG_C$ \\ \hline\hline \\[-2ex]
Cut system $\CC$ + trivalent
 &  Lagrangian description with \\[1ex]
graph $\Gamma$ on $C,$ $\si=(\,\CC\,,\,\Ga\,)$ & action functional $S^\si_\tau$
\\[2ex] \hline \\[-2ex]
cutting curve $\ga_e$ & vector multiplet $(A_{e,\mu},\phi_{e},\dots)$ \\[1ex]
$n$ boundaries  & $n$ hypermultiplets 
\\[1ex] \hline \\[-2ex]
Gluing parameters $q_e=e^{2\pi i \tau_e}$,
& UV-couplings $\tau=(\tau_1,\dots,\tau_h)$, \\
$e=1,\dots,h$, $h:=3g-3+n$ & $\displaystyle{\tau_e=\frac{4\pi i}{g_e^2}+\frac{\theta_e}{2\pi}}$
\\[2ex] \hline \\[-2ex]
Change of pants decomposition & various dualities \\[1ex] \hline
\end{tabular}
\end{center}

\subsection{Seiberg-Witten theory}

The low-energy effective actions of class $\CS$ theories are
determined as follows. Given a quadratic differential $t$ on $C$ one
defines the Seiberg-Witten curve $\Sigma_{\rm \sst SW}$ in $T^*C$ as
follows:
\begin{equation}\
\Sigma_{\rm \sst SW}\,=\,
\big\{\,(u,v)\in T^*C\,;\,v^2+t(u)=0\,\big\}\,.
\end{equation}
The curve $\Sigma_{\rm \sst SW}$ is a two-sheeted covering of $C$ with genus $4g-3+n$.
One may embed the Jacobian of $C$ into the Jacobian of $\Sigma_{\rm \sst SW}$ by
pulling back  the holomorphic differentials on $C$  under the projection $\Sigma_{\rm \sst SW}\ra C$.
Let $H_1'(\Sigma_{\rm \sst SW},\BZ)=H_1(\Sigma_{\rm \sst SW},\BZ)/H_1(C,\BZ)$, and
let us introduce a canonical basis $\BB$ for
$H_1'(\Sigma_{\rm \sst SW},\BZ)$, represented by a collection of
curves $(\al_1,\dots,\al_h;\al_1^{\rm\sst D},\dots,\al_h^{\rm\sst D})$
with intersection index $\al_k\circ \al_l^{\rm\sst D}=\de_{kl}$,
$\al_k\circ \al_l=0$, $\al_k^{\rm\sst D}\circ \al_l^{\rm\sst D}=0$.
The corresponding periods of the canonical differential on $v=v(u)du$
are defined as
\begin{equation}
a_k\,=\,\int_{\al_k}v\,,\qquad
a_k^{\rm\sst D}\,=\,\int_{\al_k^{\rm\sst D}}v\,.
\end{equation}
Using the Riemann bilinear relations,
it can be shown that there exists a function $\CF(a)$, $a=(a_1,\dots,a_h)$
such that $a_k^{\rm\sst D}=\partial_{a_k}\CF(a)$. The function $\CF(a)$ is
the prepotential determining the low-energy effective action
associated to $\BB$.

Different canonical bases $\BB$ for $H_1'(\Sigma_{\rm \sst SW},\BZ)$
are related by $Sp(2h,\BZ)$-transformations describing
electric-magnetic dualities in the low-energy physics. It will be
useful to note that for given data $\si$ specifying UV-actions there
exists a preferred class of bases $\BB_\si$ for $H_1'(\Sigma_{\rm \sst SW},\BZ)$ which are
such that the curves $\al_e$ project to the curves $\ga_e\in\CC$,
$e=1,\dots,h$ defining the pants decomposition $\CC$,
respectively.

\subsection{Relation to the Hitchin system}
\label{hit1}

The Seiberg-Witten analysis of the theories $\CG_C$ has a well-known
relation to the mathematics of the Hitchin system
\cite{Hitchin:1986vp,Hitchin:1987mz} that we will recall next.

The phase space $\CM_{\rm \sst H}(C)$ of the Hitchin system for
$G=SL(2)$ is the moduli space of pairs $(\CE,\vf)$, where $\CE$ is a
holomorphic rank 2 vector bundle with fixed determinant, and $\vf\in H^0(C,{\rm End}(\CE)\otimes K_C)$
is called the Higgs field.
The complete integrability of the Hitchin system is demonstrated using
the so-called Hitchin map. Given a pair $(\CE,\vf)$, we define the spectral curve $\Sigma$ as
\begin{equation}\
\Sigma\,=\,\big\{\,(u,v)\in T^*C\,;\,2v^2={\rm tr}(\vf^2(u))\,\big\}\,.
\end{equation}
To each pair $(\CE,\vf)$ one associates a line bundle $L$ on $\Sigma$,
the bundle of eigenlines of $\vf$ for a given eigenvalue
$v$. Conversely, given a pair $(\Sigma,L)$, where $\Sigma\subset T^*C$
is a double cover of $C$, and $L$ a holomorphic line bundle on
$\Sigma$, one can recover $(\CE,\vf)$ via
\begin{equation}
(\CE,\vf)\,:=\,\big(\,\pi_*(L)\,,\,
\pi_*(v)\,\big)\,,
\end{equation}
where $\pi$ is the covering map $\Sigma\ra C$, and $\pi_*$ is the direct
image.

The spectral curves $\Sigma$ can be identified with the curves
$\Sigma_{\rm\sst SW}$ determining the low-energy physics of the
theories $\CG_C$ on ${\mathbb R}^4$.  However, in order to give
physical meaning to the full Hitchin system one needs to consider an
extended set-up.  One possibility is to introduce surface operators.

\subsection{Two types of surface operators}

When the 6d fivebrane world-volume is of the form $M_4 \times C$,
where $C$ is a Riemann surface, there are two natural ways to
construct half-BPS surface operators in the four-dimensional
space-time $M_4$ where the $\CN=2$ theory $\CG_C$ lives.  First, one
can consider codimension-2 defects supported on $D \times C$, where $D
\subset M_4$ is a two-dimensional surface (= support of a {\it
  surface} operator).  Another, seemingly different way, is to start
with codimension-4 defects supported on $D \times \{ p \}$, where $p
\in C$ is a point on the Riemann surface.

In the case of genus-1 Riemann surface $C=T^2$, both types of half-BPS
surface operators that we study in this paper were originally
constructed using branes in \cite{Gukov:2006jk,Gukov:2008sn}.
In these papers it was argued that the two types of operators are
equivalent, at least for certain ``supersymmetric questions''.
Here we will show that for more general Riemann surfaces $C$ the two
surface operators, based on codimension-4 and codimension-2 defects,
may be different in the UV but become essentially the same in the IR
regime.  They correspond to two different ways to describe the {\em
  same} physical object.  Mathematically, this duality of descriptions
corresponds to the possibility of choosing different coordinates on
the Hitchin moduli space, which will be introduced shortly.  At first,
the equivalence of the two types of surface operators may seem rather
surprising since it is not even clear from the outset that they
preserve the same subalgebra of the supersymmetry algebra.  Moreover,
the moduli spaces parametrizing these surface operators appear to be
different.

Indeed, one of these moduli spaces parametrizes collections of $n$
codimension-4 defects supported at $D \times \{ p_i\} \subset M_4
\times C$, and therefore it is \be \text{Sym}^n (C) := C^n / S_n
\label{SymC}
\ee (Here we consider only the ``intrinsic'' parameters of the surface
operator, and not the position of $D \subset M_4$, which is assumed to
be fixed.) On the other hand, a surface operator constructed from a
codimension-2 defect clearly does not depend on these parameters,
since it wraps on all of $C$. Instead, a codimension-2 surface
operator carries a global symmetry $G$ --- which plays an important
role {\it e.g.} in describing charged matter --- and, as a result, its
moduli space is the moduli of $G$-bundles on $C$, \be \text{Bun}_G (C)
\label{BunGC}
\ee Therefore, it appears that in order to relate the two
constructions of surface operators, one must have a map between
\eqref{SymC} and \eqref{BunGC}: \be
\label{xtoumap}
\begin{array}{ccc}
\text{Bun}_G (C) & \longrightarrow & \text{Sym}^n (C) \\
x & \mapsto & u
\end{array}
\ee
where $n = (g-1) \dim G = \dim \text{Bun}_G (C)$.

It turns out that even though such a map does not exist, for $G=SL(2)$
there is a map of the corresponding cotangent bundles, which is
sufficient for our purposes. This is the celebrated classical {\em separation of variables}.
Moreover, it has a quantum version, described in Section \ref{Liou-WZW}. The separation of variables allows
us to identify the 4d theories with two types of surface operators in the IR.

The unbroken SUSY makes it possible to turn on an Omega-deformation, allowing us to define
generalizations of the instanton partition functions. In the case of codimension-2 surface operators it
turned out that the generalized instanton partition functions are calculable by the localization method,
and in a few simple cases it was observed that the results are related to the conformal blocks in the $SL(2)$-WZW model.
For codimension-4 surface operators one expects to find a similar relation to
Liouville conformal blocks with a certain number of degenerate fields inserted.


\section{Surface operators corresponding to the codimension-2 defects}
\label{sec:codim2}


Our goal in this paper is to establish a relation between the surface
operators constructed from codimension-2 and codimension-4 defects.\footnote{Even though our main examples will be
theories of class $\CS$, we expected our results --- in particular, the IR duality --- to hold more generally.}
In order to do that, we must show that they preserve the same subalgebra
of the supersymmetry algebra. This will be achieved by realizing these
defects using branes in M-theory (as we already mentioned earlier).
This realization will enable us to link the two types of
defects, and it will also illuminate their features.

In this section we present an M-theory brane construction of the
codimension-2 defects and then discuss them from the point of view of
the 4d and 2d theories. Then, in Section \ref{codim4}, we will deform
--- in a way that manifestly preserves supersymmetry --- a brane system
that gives rise to the codimension-2 defects into a brane system that
gives rise to codimension-4 defects. Using this deformation, we will
show that the two types of defects indeed preserve the same
supersymmetry algebra, and furthermore, we will connect the two types
of defects, and the corresponding 4d surface operators, to each other.

\subsection{Brane construction}
\label{codim2-1}

Following \cite{Gukov:2006jk}, we denote the support (resp. the fiber
of the normal bundle) of the surface operator inside $M_4$ by $D$
(resp. $D'$).  In fact, for the purposes of this section, we simply
take $M_4 = D \times D'$.  Our starting point is the following ``brane
construction'' of 4d $\CN=2$ gauge theory with a half-BPS surface
operator supported on $D \subset M_4$ ($= D \times D'$):
\begin{align}\label{M5M5}
\hbox{M5} &: \quad D \times D' \times C \cr
\hbox{M5$'$}  &: \quad D \times C \times D''
\end{align}
embedded in the eleven-dimensional space-time $D \times D' \times T^*
C \times \R \times D''$ in a natural way.  For simplicity, we will
assume that $D \cong D' \cong D'' \cong \R^2$ and $C$ is the only
topologically non-trivial Riemann surface in the problem at hand.
And, following the standard conventions of brane constructions
\cite{Witten:1997sc}, we use the following local coordinates on
various factors of the eleven-dimensional space-time:
\be
\label{xxxconventions}
\begin{array}{c|c|c|c|c}
D & D' & T^* C & \R & D'' \\
\hline
~x^0,~x^1~ & ~x^2,~x^3~ & ~x^4,~x^5,~x^6,~x^{10}~ & ~x^7~ & ~x^8,~x^9~ \\
\end{array}
\ee With these conventions, the brane configuration \eqref{M5M5} may
be equivalently summarized in the following diagram:
\begin{center}
\begin{tabular}{l || c|c|c|c|c|c|c|c|c|c|c}
Brane & 0 & 1 & 2 & 3 & 4 & 5 & 6 & 7 & 8 & 9 & 10 \\ \hline\hline
$M5$  & x & x & x & x &   &   & x &   &   &   & x  \\
$M5'$ & x & x &   &   &   &   & x &   & x & x & x
\end{tabular}
\end{center}
Note that $M5'$-branes wrap the same UV curve $C$ as the $M5$-branes.
This brane configuration is $\frac{1}{8}$-BPS, {\it i.e.} it preserves
four real supercharges out of 32.  Namely, the eleven-dimensional
space-time (without any fivebranes) breaks half of supersymmetry
(since $T^*C$ is a manifold with $SU(2)$ holonomy), and then each set
of fivebranes breaks it further by a half.

In particular, thinking of $T^*C$ as a non-compact Calabi-Yau 2-fold
makes it clear that certain aspects of the system \eqref{M5M5}, such
as the subalgebra of the supersymmetry algebra preserved by this
system, are not sensitive to the details of the support of M5 and
M5$'$ branes within $T^* C$ as long as both are special Lagrangian
with respect to the same K\"ahler form $\omega$ and the holomorphic
2-form $\Omega$.  Since $T^*C$ is hyper-K\"ahler, it comes equipped
with a sphere worth of complex structures, which are linear
combinations of $I$, $J$, $K$, and the corresponding K\"ahler forms
$\omega_I$, $\omega_J$, $\omega_K$.  Without loss of generality, we
can choose $\omega = \omega_I$ and $\Omega = \omega_J + i \omega_K$.
Then, the special Lagrangian condition means that both $\omega_I$ and
$\omega_K$ vanish when restricted to the world-volume of M5 and M5$'$
branes.

\subsection{Four-dimensional description}
\label{4dcodim2}

As we explain below, surface operators originating from codimension-4
defects in 6d $(0,2)$ theory naturally lead to the coupled 2d-4d
system, while those originating from codimension-2 defects in 6d
descend to the second description of surface operators in 4d gauge
theory, namely as singularities for the UV gauge fields
$A_{\mu}^{(r)}$ (see Appendix \ref{Nahm} for more details):
\begin{equation}\label{A-sing}
A_{\mu}^{(r)}dx^\mu\,\sim \,\bigg(\begin{matrix}
\chi^{(r)} & 0 \\ 0 & - \chi^{(r)} \end{matrix}\bigg)d\theta_2\,.
\end{equation}
Here, following our conventions \eqref{xxxconventions}, we use a local
complex coordinate $x^2+ix^3=r_2e^{i\theta_2}$ on $D'$ such that
surface operator is located at the origin ($r_2=0$).  A surface operator
defined this way breaks half of supersymmetry and also breaks $SO(4)$
rotation symmetry down to $SO(2) \times SO(2)$.  {}From the viewpoint
of the 2d theory on $D$, the unbroken supersymmetry is $\CN=(2,2)$.

The symmetries preserved by such a surface operator are exactly what
one needs in order to put the 4d gauge theory in a non-trivial
Omega-background. Mathematically, this leads to an $SO(2) \times
SO(2)$ equivariant counting of instantons with a ramification along
$D$.  The resulting instanton partition function
\begin{equation}
\CZ^{\rm \sst M5}_{}(a,x,\tau;\ep_1,\ep_2)\,,
\end{equation}
depends on variables $x=(x_1,\dots,x_{h})$ related to the
parameters $\chi^{(r)}$ in \rf{A-sing} via the exponentiation map \be
x_r=e^{2\pi i \tau_r\chi^{(r)}}\,. \ee
The relation between the
parameters $\chi^{(r)}$ and the counting parameters $x_r$
appearing in the instanton partition functions $\CZ^{\rm \sst M5}_{}$
was found in \cite{AT}.

\subsection{Relation to conformal field theory}

Starting from the groundbreaking work of A. Braverman \cite{Brav}, a
number of recent studies have produced evidence of relations between
instanton partition functions in the presence of surface operators
$\CZ^{\rm \sst M5}_{}(a,x,\tau;\ep_1,\ep_2)$ and conformal blocks of
affine Kac-Moody algebras $\ghat_k$ \cite{AT,KPPW,Negut,Nawata}.
Such relations can be viewed as natural generalizations of the
AGT correspondence.  In the case of class $\CS$-theories of type $A_1$
one needs to choose $\gg={\fsl}_{2}$ and $k=-2-\frac{\ep_2}{\ep_1}$,
as will be assumed in what follows.

The Lie algebra $\ghat_k$ has generators $J_n^a$, $a=0,+,-$,
$n\in\BZ$. A large class of representation of $\ghat_k$ is defined by
starting from a representation $R_j$ of the zero mode subalgebra
generated from $J_0^a$, which has Casimir eigenvalue
parametrized as $j(j+1)$.  One may then construct a representation
$\CR_j$ of $\ghat_{k}$ as the representation induced from $\CR_j$
extended to the Lie subalgebra generated by $J_n^a, n\geq 0$, such
that all vectors $v\in R_j\subset \CR_j$ satisfy $J_n^a v = 0$ for
$n>0$. To be specific, we shall mostly discuss in the following the
case that the representations $\CR_j$ have a lowest weight vector
$e_j$, but more general representations may also be considered, and
may be of interest in this context \cite{Teschner:1999ug}.

In order to define the space of conformal blocks, let $C$ be a compact
Riemann surface and $z_1,\ldots,z_n$ an $n$-tuple of points of $C$
with local coordinates $t_1,\ldots,t_n$. We attach representations
$\CR_r\equiv \CR_{j_r}$ of the affine Kac--Moody algebra $\ghat_k$ of
level $k$ to the points $z_r$, $r=1,\dots,n$.  The diagonal central
extension of the direct sum $\bigoplus_{r=1}^n \gg \otimes
\C(\!(t_r)\!)$ acts on the tensor product $\bigotimes_{r=1}^n
\CR_{r}$. Consider the Lie algebra
$$
\gg_{\out} = \gg \otimes \C[C \bsl \{ z_1,\ldots,z_n \}]
$$
of $\gg$-valued meromorphic functions on $C$ with poles allowed only
at the points $z_1,\ldots,z_n$. We have an embedding
\begin{equation}\label{gg-emb}
\gg_{\out} \,\hookrightarrow \,\bigoplus_{r=1}^n \gg
\otimes \C(\!(t_r)\!).
\end{equation}
It follows from the commutation relations in $\ghat$ and the
residue theorem that this embedding lifts to the diagonal central
extension of $\bigoplus_{r=1}^n \gg \otimes \C(\!(t_r)\!)$. Hence the
Lie algebra $\gg_{\out}$ acts on $\bigotimes_{r=1}^n \CR_{r}$.
By definition, the corresponding space of {\em conformal blocks} is
the space ${\rm CB}_{\gg}(\CR_{1},\ldots,\CR_n)$ of linear functionals
$$
\varphi: \CR_{[n]}:=\bigotimes_{r=1}^n \CR_{r} \to \C
$$
invariant under $\gg_{\out}$, i.e., such that
\begin{equation}    \label{Ward}
\varphi \left( \eta \cdot v \right) = 0, \qquad
\forall v \in \bigotimes_{r=1}^n \CR_{r}, \quad \eta \in \gg \otimes
\C[C \bsl \{ z_1,\ldots,z_n \}].
\end{equation}
The conditions \rf{Ward} represent a reformulation of current algebra
Ward identities well-known in the physics literature.  The space ${\rm
  CB}_{\gg}(\CR_{1},\ldots,\CR_n)$ is infinite-dimensional in
general.

To each $\varphi\in{\rm CB}_{\gg}(\CR_{1},\ldots,\CR_n)$ we may associate
a chiral partition function $\CZ(\varphi,C)$ by evaluating $\varphi$
on the product of the lowest weight vectors,
\begin{equation}
\CZ^{\rm \sst WZ}(\varphi,C;k):=
\varphi(e_1\otimes\ldots\otimes e_n)\,.
\end{equation}
In the physics literature one usually identifies the chiral partition
functions with expectation values of chiral primary fields
$\Phi_r(z_r)$, 
inserted at the points $z_r$,
\begin{equation}\CZ^{\rm \sst WZ}(\varphi,C;k)\,\equiv\,
\big\langle\,\Phi_n(z_n)\cdots\Phi_1(z_1)\,\big\rangle_{C,\vf}\,.
\end{equation}
Considering families of Riemann surfaces $C_\tau$ parametrized by
local coordinates $\tau$ for the Teichm\"uller space $\CT_{g,n}$ one
may regard the chiral partition functions as functions of $\tau$,
$$\CZ^{\rm \sst WZ}(\varphi,C_{\tau};k)\equiv \CZ^{\rm \sst
  WZ}(\varphi,\tau;k).$$

Large families of conformal blocks and the corresponding chiral
partition functions can be constructed by the gluing construction.
Given a (possibly disconnected) Riemann surface $C$ with two marked
points $P_0^i$, $i=1,2$ surrounded by parametrized discs
$\mathbb{D}_i$ one can construct a new Riemann surface by pairwise
identifying the points in annuli $\mathbb{A}_i\subset\mathbb{D}_i$
around the two marked points, respectively.  Assume we are given
conformal blocks $\vf_{C_i}^{}$ associated to two surfaces $C_i$ with
$n_i+1$ punctures $P_0^i,P_1^i,\dots,P_{n_i}^i$ with the same
representation $\CR_0$ associated to $P_0^i$ for $i=1,2$. Using this
input one may construct a conformal block $\vf_{C_{12}}^{}$ associated
to the surface $C_{12}$ obtained by gluing the annular neighborhoods
$\mathbb{A}_i$ of $P_0^i$, $i=1,2$ as follows:
\begin{equation}\label{gluing}\begin{aligned}
\vf_{C_{12}}^{}(v_{1}\otimes\dots\otimes & v_{n_1}\otimes
w_1\otimes\dots\otimes w_{n_2})
=\\
=&\sum_{\nu\in\mathcal{I}_{\CR_0}}\vf_{C_1}^{}(v_{1}\otimes\dots\otimes
v_{n_1}\otimes v_\nu)
\,\vf_{C_2}^{}(\CK(\tau,x)v_\nu^{\vee}\otimes w_{1}\otimes\dots\otimes
w_{n_2})\,.
\end{aligned}\end{equation}
The vectors $v_\nu$ and $v_\nu^{\vee}$ are elements of bases for the
representation $\CR_0$ which are dual w.r.t. to the invariant bilinear
form on $\CR_0$. A standard choice for the twist element
$\CK(\tau,x)\in{\rm End}(\CR_0)$ appearing in this construction is
$\CK(\tau,x)=e^{2\pi i \tau L_0}x^{J_0^0}$, where the operator $L_0$
represents the zero mode of the energy-momentum tensor constructed
from the generators $J_n^a$ using the Sugawara construction. The
parameter $q\equiv e^{2\pi i\tau}$ in (\ref{gluing}) can be identified
with the modulus of the annular regions used in the gluing
construction of $C_{12}$.  However, it is possible
to consider twist elements $\CK(\tau,x)$ constructed out a larger subset of
the generators of $\ghat_k$.  The
rest of the notation in (\ref{gluing}) is
self-explanatory. The case that $P_0^i$, $i=1,2$ are on a connected
surface can be treated in a similar way.

A general Riemann surface $C_{g,n}$ can be obtained by gluing $2g-2+n$
pairs of pants $C_{0,3}^v$, $v=1,\dots,2g-2+n$.  It is possible to
construct conformal blocks for the resulting Riemann surface from the
conformal blocks associated to the pairs of pants $C_{0,3}^v$ by
recursive use of the gluing construction outlined above. This yields
families $\vf_{j,x}^\si$ of conformal blocks parametrized by
\begin{itemize}
\item the choice of a refined pants decomposition $\si=(\CC,\Ga)$,
\item the choice of representation $\CR_{j_e}$ for each of the cutting
  curves $\ga_e$ defined by the pants decomposition, and
\item the collection of the parameters $x_{e}$ introduced via
  (\ref{gluing}) for each curve $\ga_e\in\CC$.
\end{itemize}
The corresponding chiral partition functions are therefore functions
$$\CZ^{\rm \sst WZ}_\si(j,x,\tau;k)\equiv\CZ^{\rm \sst
  WZ}(\varphi_{j,x}^\si,\tau;k).
$$
The variables $x=(x_1,\dots,x_{3g-3+n})$ have a geometric
interpretation as parameters for families of holomorphic
$G=SL(2)$-bundles $\CB$. Indeed, in  Appendix \ref{twistapp} it is explained how
the definition of the
conformal blocks can be modified in a way that depends on
the choice of a holomorphic bundle $\CB$, and why the effect of this modification can be described
using the twist elements  $\CK(\tau,x)$ appearing in the gluing construction. It follows from the
discussion in Appendix \ref{twistapp}
that changing the twist elements $\CK(\tau,x)$ amounts to
a change of local coordinates $(\tau,x)$ for the fibration of ${\rm
  Bun}_G$  over $\CT_{g,n}$ (the moduli space of pairs: a Riemann surface and a
$G$-bundle on it).

The chiral partition functions satisfy the
Knizhnik-Zamolodchikov-Bernard (KZB) equations. This is a system of
partial differential equations of the form
\begin{equation}\label{KZ}
-\frac{\ep_2}{\ep_1}
\frac{\pa}{\pa q_e}\CZ^{\rm\sst WZ}_\si(j,x,\tau;k) = \SH_e\,\CZ^{\rm\sst
  WZ}_\si(j,x,\tau;k)\,,
\end{equation}
where $\SH_e$ is a second order differential operator containing only
derivatives with respect to the variables $x_e$.  These equations can
be used to generate the expansion of $\CZ^{\rm WZ}_\si(j,x;\tau;k)$ in
powers of $q_e$ and $x_e$,
\begin{equation}\label{expansion}
\CZ^{\rm\sst WZ}_\si(j,x,\tau;k)\,\simeq\,
\sum_{\mathbf{n}\in\BZ_+^{h}}\sum_{\mathbf{m}\in\BZ_+^{h}}
\CZ^{\rm\sst WZ}_\si(j,\mathbf{m},\mathbf{n};k)\prod_{e=1}^{h}q^{\De_e+n_e}x^{j_e+m_e}\,.
\end{equation}
The notation $\simeq$ used in \rf{expansion} indicates equality up to a factor which is $j$-independent.
Such factors will be not be of interest for us.
The equations \rf{KZ} determine $\CZ^{\rm\sst
  WZ}_\si(j,\mathbf{m},\mathbf{n};k)$ uniquely in terms of $\CZ_{0,\si}^{\rm\sst
  WZ}(j)=\CZ^{\rm\sst WZ}(j,0,0;k)$. It is natural to assume that the
normalization factor $\CZ_0^{\rm\sst WZ}(j)$ can be represented as
product over factors depending on the choices of representations
associated to the three-holed spheres $C_{0,3}^v$ appearing in the
pants decomposition.

We are now going to propose the following conjecture: There exists a
choice of twist elements $\CK_e(\tau_e,x_e)$ such that we
have
\begin{equation}\label{genAGT1}
\CZ^{\rm \sst M5}_{\si}(a,x,\tau;\ep_1,\ep_2)\,\simeq\,\CZ^{\rm\sst WZ}_{\si}(j,x,q;k)\,,
\end{equation}
assuming that
\begin{equation}
j_e\,=\,-\frac{1}{2}+i\frac{a_e}{\ep_1}\,, \qquad k+2\,=\,-\frac{\ep_2}{\ep_1}\,.
\end{equation}
Evidence for this conjecture is provided by the computations performed
in \cite{AT,KPPW,Negut,Nawata} in the cases $C=C_{1,1}$ and
$C=C_{0,4}$. The relevant twist elements $\CK(\tau,x)$ were determined
explicitly in these references.
As indicated by the notation $\simeq$, we expect \rf{genAGT1} to hold only up to
$j$-independent multiplicative factors. A change of the renormalization scheme used to
define the gauge theory under consideration may modify $\CZ^{\rm \sst M5}_{}$
by factors that do not depend on $j$. Such factors are physically irrelevant,
see e.g. \cite{TV13} for a discussion.

\subsection{Relation to the Hitchin system}    \label{Hit2}

On physical grounds we expect that the instanton partition functions
$\CZ^{\rm \sst M5}_{\si}(a,x,\tau;\ep_1,\ep_2)$ behave in the limit
$\ep_1\ra 0$, $\ep_2\ra 0$ as
\begin{equation}\label{epto0}
\log\CZ^{\rm \sst M5}_{\si}(a,x,\tau;\ep_1,\ep_2)\,\sim\,-\frac{1}{\ep_1\ep_2}
\CF_\si(a,\tau)-\frac{1}{\ep_1}\tilde{\CW}^{{\rm\sst M5}}_{\si}(a,x,\tau)\,.
\end{equation}
The first term is the bulk free energy, proportional to the
prepotential $\CF_\si(a)$ defined previously. The second term is a contribution
diverging with the area of the plane on which the surface operator is
localized. It can be identified as the effective twisted superpotential
of the degrees of freedom localized on the surface $x_2=x_3=0$.

The expression of the instanton partition function as a to conformal field theory \rf{genAGT1} allows us to
demonstrate that we indeed have an asymptotic behavior of the form
\rf{epto0}. The derivation of \rf{epto0} described in Appendix
\ref{Classlim} leads to a precise mathematical description of the
functions $\tilde{\CW}^{{\rm\sst M5}}_{\si}(a,x,\tau)$ appearing in \rf{epto0}
in terms the Hitchin integrable system that we will describe in the
rest of this subsection. It turns out that $\tilde{\CW}^{{\rm\sst M5}}_{\si}(a,x,\tau)$ can be characterized as the generating
function for the change of variables between two sets of
Darboux coordinates for $\CM_{\rm \sst H}(C)$ naturally adapted to the
description in terms of Higgs pairs $(\CE,\vf)$ and pairs
$(\Sigma,L)$, respectively.

Let us pick coordinates $x=(x_1,\dots,x_{h})$ for ${\rm Bun}_{G}$.
Possible ways of doing this are briefly described in Appendix
\ref{cplxDarboux}.  One can always find coordinates $p$ on $\CM_{\rm \sst H}(C)$
which supplement the coordinates $x$ to a system of
Darboux coordinates $(x,p)$ for $\CM_{\rm \sst H}(C)$.

There exists other natural systems $(a,t)$ of coordinates for
$\CM_{\rm \sst H}(C)$ called action-angle coordinates making the
complete integrability of $\CM_{\rm \sst H}(C)$ manifest. The
coordinates $a=(a_1,\dots,a_{h})$ are defined as periods of the
Seiberg-Witten differential, as described previously.  The coordinates
$t=(t_1,\dots,t_{h})$ are complex coordinates for the Jacobian of
$\Sigma$ parametrizing the choices of line bundles $L$ on
$\Sigma$. The coordinates $t$ may be chosen such that $(a,t)$
furnishes a system of Darboux coordinates for $\CM_{\rm \sst H}(C)$.

As the coordinates $(a,t)$ are naturally associated to the description in terms
of pairs $(\Sigma,L)$, one may construct
the change of coordinates between the sets of Darboux coordinates
$(x,p)$ and $(a,t)$ using Hitchin's map introduced in Section \ref{hit1}.
The function $\tilde{\CW}^{\rm\sst
  M5}_{\si}(a,x,\tau)$ in \rf{epto0} can then be characterized as the
generating function for the change of coordinates
$(x,p)\leftrightarrow (a,t)$,
\begin{equation}\label{genfct}
  p_r=-\frac{\partial}{\pa{x_r}}\tilde{\CW}^{{\rm\sst M5}}_{\si}\,,\qquad
  t_r=\frac{1}{2\pi}\frac{\partial}{\partial{a_r}}
  \tilde{\CW}^{{\rm\sst M5}}_{\si}\,,
\end{equation}
with periods $a$ defined using a basis $\BB_\si$ corresponding to the
pants decomposition $\si$ used to define $\CZ^{\rm \sst M5}_{\si}(a,x,\tau;\ep_1,\ep_2)$.  Having defined $(x,p)$ and
$(a,t)$, the equations \rf{genfct} define $\tilde{\CW}^{{\rm\sst M5}}_{\si}(a,x,\tau)$ up to an (inessential) additive constant.

\subsection{Physical interpretation}

All of the integrable system gadgets introduced above seem to find natural
homes in field theory and string theory.  In particular, $N$
five-branes on $C$ describe a theory that in the IR corresponds to an
M5-brane wrapped $N$ times on $C$ or, equivalently, wrapped on a
$N$-fold cover $\Sigma \to C$.

Though in this paper we mostly consider the case $N=2$ (hence a double
cover $\Sigma \to C$), certain aspects have straightforward
generalization to higher ranks.  It is also worth noting that we treat
both $SL(N)$ and $GL(N)$ cases in parallel; the difference between the
two is accounted for by the ``center-of-mass'' tensor multiplet in 6d
$(0,2)$ theory on the five-brane world-volume.

Besides the ``brane constructions'' used in most of this paper, the
physics of 4d $\CN=2$ theories can be also described by
compactification of type IIA or type IIB string theory on a local
Calabi-Yau 3-fold geometry.  This approach, known as ``geometric
engineering'' \cite{Katz:1996fh,Katz:1997eq}, can be especially useful for understanding
certain aspects of surface operators and is related to the brane
construction by a sequence of various dualities.  Thus, a single
five-brane wrapped on $\Sigma \subset T^*C$ that describes the IR
physics of 4d $\CN=2$ theory is dual to type IIB string theory on a
local CY 3-fold \be z w - P(u,v) \,=\,0\,,
\label{CY3fold}
\ee
where $P(u,v)$ is the polynomial that defines the Seiberg-Witten curve
$\Sigma_{\rm\sst SW}$.

It can be obtained from our original M5-brane on $\Sigma$ by first
reducing on one of the dimensions transversal to the five-brane (down
to type IIA string theory with NS5-brane on $\Sigma$) and then
performing T-duality along one of the dimensions transversal to the
NS5-brane.  The latter is known to turn NS5-branes to pure geometry,
and supersymmetry and a few other considerations quickly tell us that
type IIB background has to be of the form \eqref{CY3fold}.

Now, let us incorporate M5$'$-brane which in the IR version of brane
configuration \eqref{M5M5} looks like:
\begin{align}\label{MMSigm}
\hbox{M5} &: \quad D \times D' \times \Sigma \cr
\hbox{M5$'$}  &: \quad D \times \Sigma \times D''
\end{align}
What becomes of the M5$'$-brane upon duality to type IIB setup
\eqref{CY3fold}?

It can become any brane of type IIB string theory supported on a
holomorphic submanifold in the local Calabi-Yau geometry
\eqref{CY3fold}. Indeed, since the chain of dualities from M-theory to
type IIB does not touch the four dimensions parametrized by $x^0,
\ldots, x^3$ the resulting type IIB configuration should still
describe a half-BPS surface operator in 4d Seiberg-Witten theory on
$M_4$.  Moreover, since type IIB string theory contains half-BPS
$p$-branes for odd values of $p$, with $(p+1)$-dimensional
world-volume, M5$'$ can become a $p$-brane supported on $D \times
C_{p-1}$, where $C_{p-1}$ is a holomorphic submanifold in a local
Calabi-Yau 3-fold \eqref{CY3fold}.

Depending on how one performs the reduction from M-theory to type IIA
string theory and then T-duality to type IIB, one finds different
$p$-brane duals of the M5$'$-brane. Here, we will be mostly interested
in the case $p=3$, which corresponds to the reduction and then
T-duality along the coordinates $x^8$ and $x^9$, {\it cf.}
\eqref{xxxconventions}.  Effectively, one can think of compactifying
the M-theory setup \eqref{MMSigm} on $D'' = T^2$, and that gives
precisely the type IIB setup \eqref{CY3fold} with extra D3-brane
supported on $\Sigma$, {\it i.e.}  at $z=w=0$ in \eqref{CY3fold}.

A D3-brane carries a rank-1 Chan-Paton bundle $L' \to \Sigma$. Therefore, we
conclude that the surface operators made from codimension-2 defects
that are obtained from the intersections with M5$'$-branes as
described above, have an equivalent description in dual type IIB
string theory in terms of pairs $(\Sigma,L')$. It seems likely that the line bundle $L'$
is closely related to the line bundle $L$ appearing in the description of the Hitchin
system in terms of pairs $(\Sigma,L)$.

Note, the degree of this line bundle, $d (L')$, is equal to the induced D1-brane charge
along the $(x^0,x^1)$ directions.
For completeness, we describe what it corresponds to in the dual M-theory setup \eqref{MMSigm}.
The T-duality that relates type IIA and type IIB brane configurations maps D1-branes
supported on $(x^0,x^1)$ into D2-branes with world-volume along $(x^0,x^1,x^8)$.
Hence, we conclude
\be
d(L') \; = \; \text{M2-brane charge along}~(x^0,x^1,x^8)
\ee
It seems worthwhile investigate the description of surface operators in terms of type IIB brane
configurations in more detail.

\section{Surface operators corresponding to codimension-4
defects}    \label{codim4}

As we mentioned earlier, there is another way to construct surface
operators in 4d $\CN=2$ theories of class ${\mathcal S}$ -- namely,
by introducing codimension-4 defects in 6d five-brane theory
\cite{Witten:1997sc,Maldacena:2000mw,Gaiotto:2008cd,Gaiotto:2009we}.

In this section we present this construction. The idea is to start
with the brane system which we used in the previous section to produce
the codimension-2 defects and to deform it in such a way that the end
result is a collection of codimension-4 defects. The advantage of this
way of constructing them is that, as we will see below, this process
does not change the subalgebra of the supersymmetry algebra preserved
by the defects. Therefore, it follows that the two types of defects in
fact preserve the same subalgebra.

In the next sections we will also use this link between the
codimension-4 and codimension-2 defects in the 6d theory in order to
establish the connection between the corresponding 4d $\CN=2$ theories
in the IR.

\subsection{Brane construction}    \label{branecon}

The origin of codimension-4 defects in 6d theory and the resulting
surface operators in 4d $\CN=2$ theory are best understood via the
following brane construction:
\begin{center}
\begin{tabular}{l || c|c|c|c|c|c|c|c|c|c|c}
Brane & 0 & 1 & 2 & 3 & 4 & 5 & 6 & 7 & 8 & 9 & 10 \\ \hline\hline
$M5$  & x & x & x & x &   &   & x &   &   &   & x  \\
$M2$ & x & x &   &   &   &   &   & x &   &   &
\end{tabular}
\end{center}
where in addition to $N$ $\; M5$-branes supported on $M_4 \times C$
(as in Section \ref{codim2-1}) we have added a number of $M2$-branes
supported on $D \times \R_+$, where $\R_+ = \{ x^7 \ge 0 \}$.  Note
that each of these $M2$-branes is localized at one point of the UV
curve $C$ and therefore gives rise to a codimension-4 defect in the 6d
theory.

One of the main goals of this paper is to show that the surface
operators in 4d $\CN=2$ theory corresponding to these codimension-4
defects describe in the IR the same physical object as \eqref{M5M5},
up to a field transformation (which is related to a change of Darboux-coordinates
in the associated integrable system). For such an
equivalence to make sense, it is necessary that the two types of
defects preserve the same supersymmetry subalgebra. This is a
non-trivial statement that we explain presently.

\FIGURE{
\includegraphics[width=5.5in]{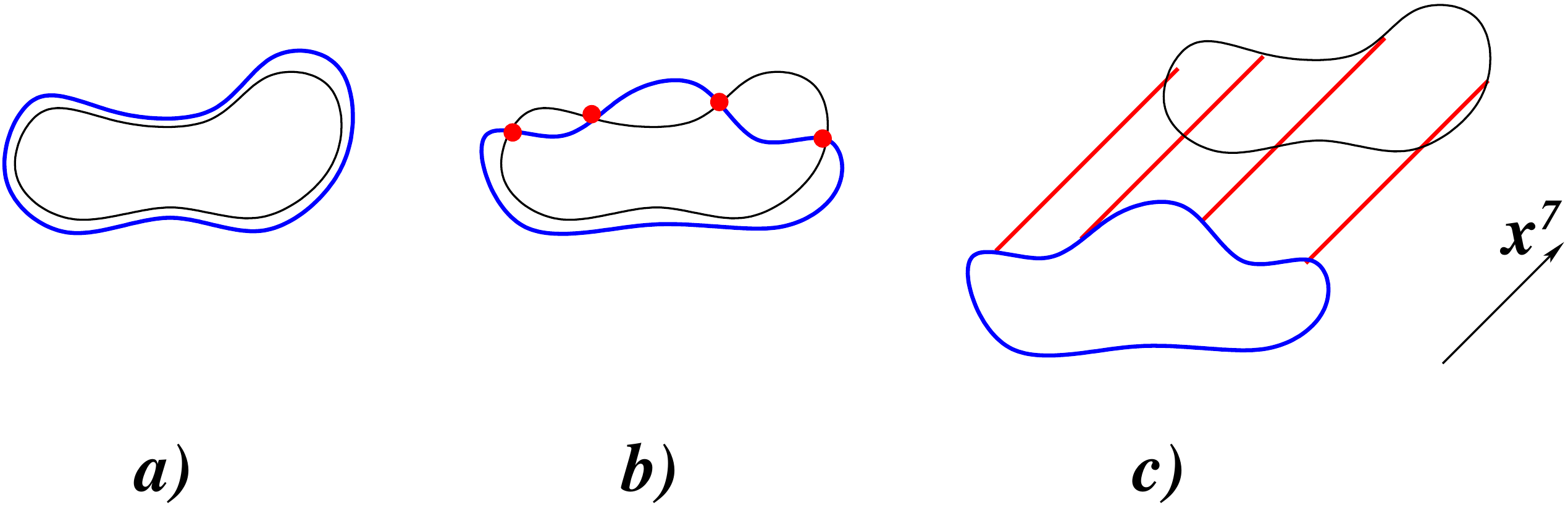}
  \caption{An M5$'$-brane wrapped on the curve $C$ can be
    perturbed to a curve $\tilde C$ which meets $C$ at finitely many
    points $u_i$. Then, separating the five-branes on $C$ and $\tilde
    C$ along the $x^7$ direction results in creation of M2-branes
    (shown in red).\label{M2creationfig}}
}

A simple and elegant way to analyze supersymmetry and to gain further
insight into the relation between the two types of surface operators
is to perform a continuous deformation of one brane configuration into
the other preserving the corresponding subalgebra of the supersymmetry
algebra.\footnote{The argument presented below applies equally well to
  a system where the UV curve $C$ is replaced by the IR curve
  $\Sigma$. In fact, the latter version, which similarly explains that
  IR surface operators preserve the same SUSY is also responsible for
  the IR duality that underlies the separation of variables map.}
Starting with our original system \eqref{M5M5}, we keep the
world-volume of the M5-branes to be $D \times D' \times C$, but deform
the support of the M5$'$-branes to be $D \times \tilde C \times D''$,
where $\tilde C \subset T^*C$ is a deformation of the zero section $C
\subset T^*C$, which is special Lagrangian with respect to $\omega =
\omega_I$ and $\Omega = \omega_J + i \omega_K$:
\begin{align}
\label{M5M5def}
\hbox{M5} &: \quad D \times D' \times C \cr
\hbox{M5$'$}  &: \quad D \quad \times \quad \tilde C \times D''
\end{align}
According to the discussion in Section \ref{codim2-1}, this
deformation does not affect the amount of unbroken supersymmetry, and
so \eqref{M5M5def} preserves the same part of the supersymmetry
algebra as the original system \eqref{M5M5}.  Note that deformations
of special Lagrangian submanifolds are infinitesimally parametrized by
$H^1 (C)$ and, in most cases of interest, this is a fairly large
space.  However, what's even more important is that, after the
deformation, $\tilde C$ meets the original curve $C$ only at finitely
many points $u_i$, as illustrated on Figure \ref{M2creationfig}$b$.
The number of such intersection points is determined by the Euler
characteristic (or genus) of the curve $C$
\begin{equation}
C \cdot C \; = \; 2 g(C) - 2 \,.
\end{equation}
At low energies one may effectively represent the stack of $M5$-branes
in terms of a smooth curve $\Sigma\subset T^*C$
\cite{Witten:1997sc}. The M5$'$-branes will be represented by a curve
$\Sigma'$ related to $\Sigma$ by holomorphic deformation.  Using the
same arguments as above one may show, first of all, that two types of
IR surface operators preserve the same SUSY and, furthermore,
determines the number of intersection points on $\Sigma$ to be
\begin{equation}
\label{Cadjunct}
\Sigma \cdot \Sigma \; = \; 2 g_\Sigma^{} - 2 \,,
\end{equation}
where $g_\Sigma^{}=4g-3$  if $C$ has no punctures \cite{Hitchin:1986vp}, as
will be assumed in this section for simplicity.

After the deformation, every intersection of M5 and M5$'$ locally
looks like a product of $\mathbb{R}^2$ with a submanifold in
$\mathbb{R}^9$, which is a union of two perpendicular 4-spaces
$\mathbb{R}^4 \cup \mathbb{R}^4$, intersecting at one point, times the
real line $\mathbb{R}$ parametrized by the coordinate $x^7$. Indeed, M5 and M5$'$
overlap along a 2-dimensional part of their world-volume, $D$, and the
remaining 4-dimensional parts of their world-volume span $\mathbb{R}^8
= \{ x^7 = 0 \}$.  If we separate these five-branes in the $x^7$
direction, they become linked in the 9-dimensional space which is the
part of the space-time orthogonal to $D$. Then, if we make one of the
five-branes pass through the other by changing the value of its
position in the $x^7$ direction, an M2-brane is created, as shown on
Figure~\ref{M2creationfig}$c$.  The support of the M2-brane is $D
\times I$, where $I$ is the interval along $x^7$ connecting the
deformations of the 4-spaces, which we denote by $\mathbb{R}^4_a$ and
$\mathbb{R}^4_b$ (where $a$ and $b$ are the values of the coordinate
$x^7$ corresponding to these two subspaces):
\begin{align}
\label{M54planes}
\hbox{M5} &: \quad D \times \mathbb{R}^4_a \cr
\hbox{M5$'$}  &: \quad D \times \mathbb{R}^4_b \cr
\hbox{M2}  &: \quad D \times \{ a \le x^7 \le b \}
\end{align}
This creation of the M2-brane between two linked M5-branes is a
variant of the so-called Hanany-Witten effect \cite{HW}. What this
means for us is that a surface operator represented by a codimension-2
defect wrapped on $D \times \Sigma$ in the fivebrane theory can be
equivalently represented by a collection of codimension-4 defects
supported at various points $u_i \in \Sigma$.

Indeed, globally, after separating M5 and M5$'$ in the $x^7$
direction, the brane configuration \eqref{M5M5def} looks like this:
\begin{align}
\label{M5M5M2C}
\hbox{M5} &: \quad D \times D' \times \Sigma \cr
\hbox{M5$'$}  &: \quad D \times D'' \times \tilde \Sigma \cr
\hbox{M2}  &: \quad D \times I
\end{align}
Here, adding M2-branes does not break supersymmetry any further, so
that \eqref{M5M5M2C} is a $\frac{1}{8}$-BPS configuration for
arbitrary special Lagrangian submanifolds $\Sigma$ and $\tilde \Sigma \subset
T^* \Sigma$.  Of course, the special case $\tilde \Sigma \equiv \Sigma$ takes us back
to the original configuration \eqref{M5M5}, schematically shown in
Figure \ref{M2creationfig}$a$.  On the other hand, separating M5 and
M5$'$ farther and farther apart, we basically end up with the standard
brane configuration, shown on Figure~\ref{branefig}$b$, that describes
half-BPS surface operator(s) built from codimension-4 defects, or
M2-branes.  In fact, even our choice of space-time conventions
\eqref{xxxconventions} agrees with the standard notations used in the
literature, so that \eqref{M5M5M2C} can be viewed as M-theory lift of
the following brane system in type IIA string theory:
\begin{align}\label{iiabranes}
\hbox{NS5} &: \quad 012345 \cr
\hbox{D4}  &: \quad 0123~~~6 \cr
\hbox{NS5$'$} &: \quad 01~~~45~~~~~89 \cr
\hbox{D2}  &: \quad 01~~~~~~~~7
\end{align}
Conversely, reduction of \eqref{M5M5M2C} on the M-theory circle
(parametrized by $x^{10}$) gives the type IIA system \eqref{iiabranes}
shown on Figure~\ref{branefig}$a$.

\FIGURE{
\includegraphics[width=5.0in]{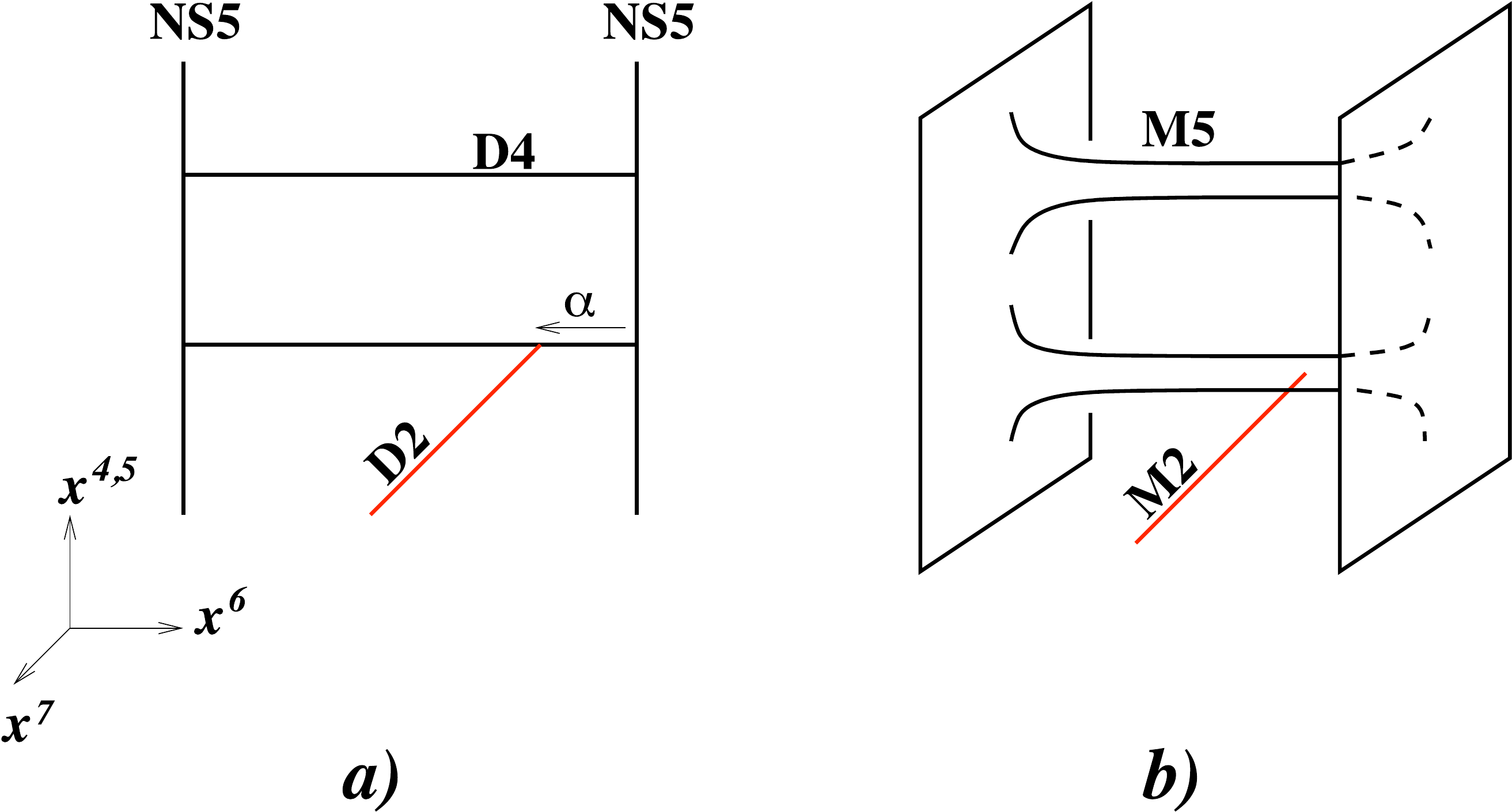} \caption{The
    brane construction of a surface operator in pure $\CN=2$ super
    Yang-Mills theory $(a)$ in type IIA string theory and $(b)$ its
    M-theory lift.
\label{branefig}}
}

How many M2-branes are created in the configuration \eqref{M5M5M2C}?
If the number of M5-branes is $N$ and the number of M5$'$-branes is
$k$, then each intersection point $u_i \in \Sigma \cap \tilde \Sigma$
contributes $k \cdot N$ M2-branes (due to the $s$-rule \cite{HW}).
When we multiply this by the number of intersection points
\eqref{Cadjunct}, we get the answer $2 (g - 1) k N$.  This number,
however, counts how many M2-branes are created as one pulls a stack of
M5$'$-branes through the stack of M5-branes by changing their
$x^7$-position from $x^7 < 0$ to $x^7 > 0$, while we are interested in
a process that starts at $x^7 = 0$ and then goes to either $x^7 < 0$
or $x^7 > 0$.

The initial value $x^7 =0$ is somewhat singular. However, as in a
similar ``geometric engineering'' of 2d field theories with the same
amount of supersymmetry \cite{Gukov:1999ya}, we shall assume that both
phases $x^7 < 0$ and $x^7 > 0$ are symmetric and the same number of
M2-branes is created (or destroyed) as we pass from $x^7 =0$ to either
$x^7 < 0$ or $x^7 > 0$. In fact, via a chain of dualities
\cite{Gukov:2002es} our ``brane engineering'' of the 2d theory on
M2-branes can be mapped to the ``geometric engineering'' of
\cite{Gukov:1999ya}, which therefore justifies applying the same
arguments. Then, it means that the answer we are looking for is only
half of $2 (g - 1) k N$, {\it i.e.}
\begin{equation}
\# \text{(M2-branes)} \; = \; (g - 1) k N
\end{equation}
The case considered in this paper is $N=k=2$, giving a number of $4g-4$ M2-branes created.

In the IR one may represent the M5 by a  curve $\Sigma$
in $T^*C$.  The M5$'$-branes are supported on a holomorphic deformation of $\Sigma$, which
may be represented by a section of a line bundle of the same degree as $K_{\Sigma}$,
\begin{equation}
{\rm deg}(K_{\Sigma})\,=\,2g_\Sigma^{}-2 \,=\,2(4g-3) -2\,=\,
8g-8\,.
\end{equation}
It seems natural to assume that $\Sigma'$ is symmetric under the involution exchanging the
two sheets of $\Sigma$. This implies that the projection
$\pi:\Sigma\ra C$ of the intersection points defines $4g-4$ points $u=(u_1,\dots,u_{4g-4})$ on $C$.
Following the discussion above, one expects to find a collection of
M2-branes created with end-points at $u_r$, $r=1,\dots,4g-4$.


Since a surface operator supported on $D \subset M_4$ breaks
translation invariance in the transverse directions (along $D'$), it
must necessarily break at least part of supersymmetry of the 4d
$\CN=2$ gauge theory on $M_4$.  In addition, our analysis above shows
that both types of surface operators preserve the same part of
supersymmetry.  It is convenient to express the unbroken parts of 4d
Lorentz symmetry and supersymmetry in 2d language.  Indeed, the
unbroken generators of the Lorentz symmetry (in $x^0$ and $x^1$
directions along $D$) conveniently combine with the unbroken
supercharges and the R-symmetry generators to form 2d $\CN=(2,2)$
supersymmetry algebra.

\subsection{Four-dimensional description}    \label{4d descr}

We now start discussing the implications of this construction for the
IR physics of 4d $\CN=2$ gauge theories with surface operators.

The Lagrangian of a 4d $\CN=2$ gauge theory with surface operators may
have additional terms corresponding to 2d $\CN=(2,2)$ supersymmetric
theories coupled to the surface operators. Recall that the Lagrangian
of a theory with 2d $\CN=(2,2)$ supersymmetry is allowed to have a
particular type of F-term called the {\em twisted superpotential},
denoted by $\tilde \CW$. From the point of view of a 4d theory, such a
term is a two-dimensional feature, {\it i.e.} such terms would not be
present in a 4d $\CN=2$ theory without surface
operators, 
and it is partially
protected by the supersymmetry from quantum corrections.
Moreover, in the IR, the 4d $\CN=2$ gauge theory with surface
operators is completely determined by the prepotential $\CF$ and the
twisted superpotential $\tilde \CW$ (see {\it e.g.} \cite{Gukov:2014gja}
for a recent review).

Recall that the low-energy effective action has a four-dimensional
part and a two-dimen\-sional part,
\begin{equation}
S\,=\,\int d^4xd^4\theta\;\CF+\left(\frac{1}{2}\int
  d^2xd^2\tilde{\theta}\;\tilde{\CW}+{\rm c.c.}\right)\,,
\end{equation}
where $\CF$ is the prepotential giving the low-energy effective action
of the four-dimensional theory in the absence of a surface operator,
and $\tilde{\CW}$ is the holomorphic twisted superpotential.  We will
mostly consider $\CF$ as a function $\CF(a,\tau)$, with $a$ being a
collection $a=(a_1,\dots,a_h)$ of coordinates for the moduli space of
vacua $\CM_\text{vac}$, where $h$ is the dimension of
$\CM_\text{vac}$, and $\tau$ being the collection of UV gauge coupling
constants $\tau=(\tau_1,\dots,\tau_h)$.  The dependence on the mass
parameters will not be made explicit in our notations.
$\tilde{\CW}\equiv \tilde{\CW}(a,\kappa,\tau)$ depends on $a$ and
$\tau$, and may furthermore depend on a collection of parameters
$\kappa$ characterizing the surface operator in the UV.

The presence of surface operators implies that the abelian gauge
fields $A_r$, $r=1,\dots,h$ appearing in the same vector-multiplet as
the scalars $a_r$ will generically be singular at the support $D$ of
the surface operator.  The singularity is such that the field strength
$F_r$ associated to $A_r$ has a singularity of the form
$(F_r)_{23}=2\pi \al_r\de(x_2)\de(x_3)$. The parameters $\al_r$ are
related to the twisted superpotential $\tilde{\CW}$ by a relation of the
form
\begin{equation}\label{tfromW}
  t_r\,\equiv\,\eta_r+\tau_{rs}\al_s:=\frac{1}{2\pi}\frac{\pa}{\pa
    a_r}\tilde{\CW}\,,\qquad\tau_{rs}:= \frac{\pa}{\pa
    a_r}\frac{\pa}{\pa a_s}\CF\,.
\end{equation}
The parameters $\eta_r$ in \rf{tfromW} characterize the divergence of
the dual gauge fields in a similar way. As indicated in \rf{tfromW},
it is useful to combine the Gukov-Witten parameters $\al_r$ and
$\eta_r$ into complex variables $t=(t_1,\dots,t_h)$ which are
functions of $a$, $\tau$ and $\kappa$.

The argument of the previous subsection shows that the brane
configuration \eqref{M5M5} that describes codimension-2 defects can be
continuously deformed without changing the unbroken supersymmetry to a
brane configuration describing codimension-4 defects:
\begin{align}
\label{M5M2}
\hbox{M5} &: \quad D \times D' \times C \cr
\hbox{M2}  &: \quad D \times \R_+
\end{align}
This has important implications for our story. First, it means that
the same type of Omega-background in both cases leads to the same kind
of F-terms (appearing in the instanton partition functions) for both
types of surface operators. Namely, in the language of unbroken 2d
$\CN=(2,2)$ supersymmetry, it is the twisted superpotential $\tilde
\CW$ in both \eqref{epto0} and \eqref{ZinstM2}.

Note that by itself, the existence of a continuous deformation
relating surface operators corresponding to the codimension-2 defects
to those corresponding to the codimension-4 defects does not
necessarily imply their equivalence.  Indeed, there are many physical
systems related by a continuous deformation which describe completely
different physics, {\it e.g.} gauge theory at different values of a
coupling constant is a simple example.  However, certain quantities
may be insensitive to a change of parameter, and in fact, in the case
at hand, we will show that the twisted superpotential $\tilde \CW$ is
precisely such a quantity that does not depend on the deformation
described in the previous subsection (up to a change of variables).

But the twisted superpotential $\tilde \CW$ determines the vacuum
structure and the IR physics of the 4d theories with surface
operators. Therefore if we can show that $\tilde \CW$ is independent
of the deformation, it will follow that the corresponding 4d theories
are equivalent in the IR.

So, our plan is the following. In this subsection, we show that the
twisted superpotential $\tilde \CW$ is indeed independent of the
separation of M5 and M5$'$ in the $x^7$ direction, which was our
deformation parameter in the brane configuration \eqref{M5M5M2C} that
interpolates between \eqref{M5M5} and \eqref{M5M2}. And then, in the
next section, we will use this independence of $\tilde \CW$ on the
deformation parameter to argue that the 4d theories with the surface
operators corresponding to the codimension-2 and codimension-4 defects
describe the same physics in the IR regime (in other words, they are
related by an IR duality).

In order to show the $x^7$-independence of $\tilde \CW$, we need to focus more
closely on the surface operators produced from codimension-4 defects
and explain a few facts about the brane systems
\eqref{M5M5M2C}--\eqref{M5M2} that involve M2-branes.  As we already
pointed out earlier, the brane configuration \eqref{M5M5M2C} is simply
an M-theory lift of the brane system \eqref{iiabranes} illustrated in
Figure \ref{branefig}$a$.  Usually, such M-theory lifts capture IR
quantum physics of the original type IIA system, {\it cf.}
\cite{Witten:1997sc}.  In the present case, the relevant theory
``lives'' on D4-branes and D2-branes in \eqref{iiabranes}.  The theory
on D4-branes is simply the 4d gauge theory on $M_4$, and describing
its IR physics via its M-theory lift was one of the main points of
\cite{Witten:1997sc}.  The theory on D2-branes is a 2d theory with
$\CN=(2,2)$ supersymmetry preserved by the system \eqref{iiabranes},
see {\it e.g.} \cite{Hanany:1997vm,Hanany:2003hp,AGGTV,DGL}.
This 2d theory couples to 4d gauge theory and, hence,
describes a half-BPS surface operator as a combined 2d-4d system.

This has to be compared with our earlier discussion in Section
\ref{4dcodim2}, where we saw that surface operators constructed from
codimension-2 defects naturally lead to singularities of gauge fields
in the 4d gauge theory, while now we see that surface operators built
from codimension-4 defects naturally lead to a description via
combined 2d-4d system.  Furthermore, the number $N$ of D4-branes that
determines the rank of the {\it gauge} group in four dimensions is the
rank of the {\it flavor} symmetry group from the viewpoint of 2d
theory on the D2-branes.  In particular, in the basic case of $N=2$
each D2-branes carries a $U(1)$ linear sigma-model with $N=2$ charged
flavors, whose Higgs branch is simply the K\"ahler quotient $\C^2
/\!\!/ U(1) \cong \BC\BP^1$.

This implies that codimension-4 defects give rise to a 2d-4d coupled
system, in which gauge theory in the bulk is coupled to the $\BC\BP^1$
2d sigma-model on $D \subset M_4$, which is IR-equivalent to the
corresponding 2d gauged linear sigma model. Moreover, this also shows
why the deformation associated to the separation along $x^7$ direction
in \eqref{M5M5M2C} does not affect the corresponding twisted
superpotential. And here the identification of unbroken supersymmetry
and the precise type of the F-terms in 2d becomes crucial.

Namely, from the viewpoint of the D2-branes in \eqref{iiabranes}, the
separation along the $x^7$ direction is the gauge coupling constant of
the 2d gauged linear sigma-model
\cite{Hanany:1997vm,Hanany:2003hp,AGGTV,
DGL},
\begin{equation}
g_{\text{2d}}
= {\Delta x^{7} \over \ell_s^2} \Bigg|_{{\rm D2}}
\end{equation}
On the other hand, it is a standard fact about 2d $\CN=(2,2)$
supersymmetry algebra that twisted superpotential is independent on
the 2d gauge coupling constant \cite{Witten:1993yc}.

The reader may observe that the number of variables $u_i$
parametrizing the positions of the created M2-branes exceeds the
number of parameters $\chi^{(r)}$ introduced via \rf{A-sing} for
surfaces of genus $g>1$. At the moment it does not seem to be known
how exactly one may describe the system with M5- and M5$'$-branes at
an intermediate energy scale in terms of a four-dimensional quantum
field theory. It seems quite possible that the resulting description
will involve coupling one gauge field $A_{\mu}^{(r)}$ to more than one
copy of the $\BC\BP^1$ 2d sigma-model on $D \subset M_4$, in general.


\subsection{Twisted superpotentials as generating
  functions}    \label{gen fn}

As we have seen in the previous subsection, regardless how different
the theories with two types of surface operators may be in the UV,
their effective descriptions in the IR have a relatively simple and
uniform description. More specifically, the theories we are
considering in this paper are essentially determined in the IR by
their twisted superpotentials. Hence we focus on them.

The twisted superpotentials in the presence of codimension-2 and
codimension-4 surface operators will be denoted by $\tilde
\CW^{{\rm\sst M5}}$ and $\tilde \CW^{\rm\sst M2}$, respectively. The
twisted superpotential $\tilde \CW^{{\rm\sst M5}}\equiv \tilde
\CW^{{\rm\sst M5}} (a,x,\tau)$ depends besides $a$ and $\tau$ on
coordinates $x$ for $\text{Bun}_G (C)$, and $\tilde \CW^{\rm\sst
  M2}\equiv\tilde \CW^{\rm\sst M2} (a,u,\tau)$ on the positions of the
points on $C$ where the codimension-2 defects are located.

From both $\tilde \CW^{{\rm\sst M5}}$ and $\tilde \CW^{\rm\sst M2}$
we can find the corresponding Gukov-Witten parameters $t^{{\rm\sst
    M5}}(a,x,\tau)$ and $t^{\rm\sst M2}(a,u,\tau)$ via
\rf{tfromW}. If the two surface operators are equivalent in the deep
IR there must in particular exist an analytic, locally invertible
change of variables $u=u_\ast(x;a,\tau)$ relating the Gukov-Witten
parameters $t$ and $t'$ as
\begin{equation}
t^{{\rm\sst M5}}(a,x,\tau)=t^{\rm\sst M2}(a,u_\ast(x;a,\tau),\tau)\,.
\end{equation}
It follows that the twisted superpotentials $\tilde{\CW}^{{\rm\sst
    M5}}$ and $\tilde{\CW}^{\rm\sst M2}$ may differ only by a
function independent of $a$.

One may furthermore note that the variables $u_i$ are dynamical at
intermediate scales, or with non-vanishing Omega-deformation. The
system obtained by separating the M5$'$-branes by some finite distance
$\Delta x^7$ from the M5-branes will be characterized by a
superpotential $\tilde{\CW}'$ depending both on $x$ and $u$, in
general. We had argued above that this superpotential does not depend on
the separation $\Delta x^7$.  Flowing deep into the IR region one
expects to reach an effective description in which extremization of
the superpotential determines $u$ as function of $x$ and the remaining
parameters, $u=u_\ast(x,a,\tau)$.  The result should coincide with
$\tilde \CW^{{\rm\sst M5}} (a,x,\tau)$, which is possible if the
resulting superpotential $\tilde{\CW}'$ differs from
$\tilde\CW^{\rm\sst M2} (a,u,\tau)$ by addition of a function
$\tilde\CW''(u,x,\tau)$ that is $a$-independent
\begin{equation}
\tilde\CW'(a,x,u,\tau)\,=\,
\tilde\CW^{\rm\sst M2} (a,u,\tau)+\tilde\CW''(u,x,\tau)\,;
\end{equation}
the additional piece $\tilde\CW''(u,x,\tau)$ may be attributed to the
process creating the M2-branes from M5$'$-branes.  Extremization of
$\tilde{\CW}'$ implies that
\begin{equation}\label{extremW}
  \frac{\pa}{\pa u_r}\tilde\CW^{\rm\sst M2}
  (a,u,\tau)\Big|_{u=u_\ast(x,a,\tau)}\,=\,
  -\frac{\pa}{\pa u_r}\tilde\CW''(u,x,\tau)\Big|_{u=u_\ast(x,a,\tau)}\,,
\end{equation}
and $\tilde{\CW}'(a,x,u,\tau)\big|_{u=u_\ast}$ should coincide with
$\tilde \CW^{{\rm\sst M5}} (a,x,\tau)$.

We are now going to argue that $\CW^{{\rm\sst M5}}$, $\CW^{\rm\sst
  M2}$ and $\tilde\CW''$ represent {\em generating functions} for
changes of variables relating three different sets of
Darboux-coordinates for the same moduli space $\CM_{\rm\sst 2d}$
locally parametrized by the variables $a$ and $x$ (see, for example,
\cite{Ramond}, Section 2.1, for the definition of generating functions
and a discussion of their role in the Lagrangian formalism).

Considering $\CW^{{\rm\sst M5}}$ first, one may define other local
coordinates for $\CM_{\rm\sst 2d}$ as
\begin{equation}\label{pfromW}
p_r\,=\,-\frac{\pa}{\pa x_r}\tilde\CW^{{\rm\sst M5}} (a,x,\tau)\,.
\end{equation}
Both $(x,p)$ and $(a,t)$, with $t$ defined via \rf{tfromW}, will
generically define local coordinates for $\CM_{\rm\sst 2d}$. Having a
Poisson-structure on $\CM_{\rm\sst 2d}$ that makes $(x,p)$ into
Darboux-coordinates it follows from \rf{tfromW} and \rf{pfromW} that
$(a,t)$ will also be Darboux-coordinates for $\CM_{\rm\sst 2d}$.

If $x$ and $u$ are related by a locally invertible change of variables
$u=u_\ast(x;a,\tau)$ it follows from \rf{extremW} that $u$ together
with the coordinates $v$ defined by
\begin{equation}
v_r\,=\,\frac{\pa}{\pa u_r}\tilde\CW^{\rm\sst M2} (a,u,\tau)\,,
\end{equation}
will represent yet another set of Darboux coordinates for
$\CM_{\rm\sst 2d}$.  In this way one may identify $\CW^{\rm\sst M2}$
and $\CW'$ as the generating functions for changes of Darboux
variables $(a,t)\leftrightarrow(u,v)$ and $(u,v)\leftrightarrow(x,p)$
for $\CM_{\rm vac}$, respectively.


There are various ways to compute the twisted superpotential $\tilde
\CW$.  One (though not the only one!) way is to compute the asymptotic
expansion of the Nekrasov partition function \cite{Nekrasov:2002qd} in
the limit $\epsilon_{1,2} \to 0$. It takes the form
\be
\label{ZinstM2}
\log Z^{\text{inst}} = -\frac{\CF}{\epsilon_1 \epsilon_2} -
\frac{\tilde \CW}{\epsilon_1} + \ldots
\ee
Here, $\CF$ is the Seiberg-Witten
prepotential that does not depend on the surface operator and defines
the corresponding IR 4d theory in the bulk. The next term in the
expansion, $\tilde \CW$, is what determines the IR theory with the
surface operator.\footnote{In a system without surface operators one has
$\tilde \CW = 0$.}

In what follows we will use the relations of the instanton partition
functions to conformal blocks to determine $\tilde \CW^{{\rm\sst M5}}
(a,x,\tau)$ and $\tilde \CW^{\rm\sst M2} (a,u,\tau)$ via
\rf{ZinstM2}. Both functions will be identified as generating
functions for changes of Darboux-variables $(x,p) \leftrightarrow
(a,t)$ and $(u,v) \leftrightarrow (a,t)$ for the Hitchin moduli space
$\CM_{\rm\sst H}(C)$, respectively. Among other things, this will
imply that $\tilde \CW^{{\rm\sst M5}} (a,x,\tau)$ and $\tilde
\CW^{\rm\sst M2} (a,m;u)$ indeed satisfy a relation of the form
\begin{equation}\label{W-rel}
\tilde \CW^{{\rm\sst M5}} (a,x,\tau)\,=\,
\tilde\CW^{\rm\sst M2} (a,u_\ast(x,a,\tau),\tau)+\tilde\CW^{\rm\sst
  SOV}(u_\ast(x,a,\tau),x,\tau)\,.
\end{equation}
In view of the discussion above one may view this result as nontrivial
support for the conjectured IR duality relation between the theories
with the surface operators of co-dimensions 2 and 4, if we set
$\tilde\CW''\equiv\tilde\CW^{\rm\sst SOV}$.

\subsection{Relation to conformal field theory}

We had previously observed that the twisted superpotentials
$\tilde{\CW}^{{\rm\sst M5}}_{\si}(a,x,\tau)$ that may be calculated
from the instanton partition functions $\CZ^{\rm \sst
  M5}_{\si}(a,x,\tau;\ep_1,\ep_2)$ via \rf{epto0} represent changes
of Darboux variables for the Hitchin integrable system. We will now
discuss analogous results for $\tilde{\CW}^{\rm\sst
  M2}_{\si}(a,u,\tau)$. To this aim we begin by describing the
expected relations between the instanton partition functions $\CZ^{\rm
  \sst M2}_{\si}(a,x,\tau;\ep_1,\ep_2)$ and Liouville conformal
blocks.

Conformal blocks for the Virasoro algebra with central charge
$c_b=1+6(b+b^{-1})^2$ may be defined in close analogy to the Kac-Moody
conformal blocks discussed above. Our discussion shall therefore be
brief.  Given a Riemann surface $C$ with $n$ punctures, we associate
representations $\CV_{\al_r}$ generated from highest weight vectors
$v_{\al_r}$ to the punctures $z_r$, $r=1,\dots,l$. The Lie algebra
$\on{Vect}(C\setminus\{z_1,\dots,z_l\})$ of meromorphic vector fields
on $C$ with poles only at $z_r$, $r=1,\dots,l$, is naturally embedded
into the direct sum of $l$ copies of the Virasoro algebra with the
central elements identified (using the expansion of the vector fields
near the punctures).  Conformal blocks $\vf$ are then defined as
linear functionals on $\bigotimes_{r=1}^l\CV_{\al_r}$ that are
invariant under the action of
$\on{Vect}(C\setminus\{z_1,\dots,z_l\})$. This invariance condition
represents the conformal Ward identities.  Chiral partition functions
$\CZ_\CF(\vf,C;b)$ are defined as the evaluation of $\vf$ on the product of
highest weight vectors $\bigotimes_{r=1}^lv_{\al_r}$, in the physics literature
often denoted as
\begin{equation}
\CZ_\CF(\vf,C;b)\,\equiv\,\big\langle\,e^{2\al_n\vf(z_n)}\cdots e^{2\al_1\vf(z_1)}\,\big\rangle_{C,\vf}\,.
\end{equation}

In general, the space of conformal blocks is
infinite-dimensional. However, it can be decomposed into a direct sum
(or direct integral, depending on the situation) of finite-dimensional
spaces (in some cases, such as that of the Liouville model,
one-dimensional spaces, so that we obtain a basis) using the gluing
construction reconstructing $C$ from its pants decompositions
specified by the data $\si=(\CC,\Ga)$ introduced in Section
\ref{classS1}. Its elements are labeled by representation parameters
$\beta_e$ assigned to the cut curves $\ga_e\in\CC$. We denote the
resulting chiral partition functions by $\CZ^{\rm\sst L}(\beta,\tau;b)$.

We shall also discuss the situation of $d$ additional degenerate
representations $\CV_{-1/2b}$ (sometimes called $\Phi_{1,2}$ primary
fields) associated to points $S=\{u_1,\dots,u_d\}\subset C$ that are
distinct and different from the punctures $z_1,\dots,z_l$. The
corresponding chiral partition functions then satisfy $d$ second order
differential equations resulting from the existence of degree 2 null
vectors in $\CV_{-1/2b}$.  A basis for the space of solutions can be
obtained by starting from a pants decomposition $\si$ of $C$. Each
pair of pants $C_{0,3}^v$ obtained by cutting along $\CC$ contains a
subset $S_v$ of $S$. Choosing a pants decomposition of
$C_{0,3}^v\setminus S_v$ one obtains a refined pants decomposition
$\hat\si$ that can be used to define chiral partition functions
$\CZ^{\rm\sst L}_{\hat\si,\varpi}(\b,u,q;b)$ as before.
The additional set of labels $\varpi$ entering the definition of $\CZ^{\rm\sst L}_{\hat\si,\varpi}$ is
constrained by the fusion rules for existence of conformal blocks with degenerate representations
inserted, and may therefore be represented by elements of
$\BZ_2^d$.

The precise definition of the instanton partition functions
$\CZ^{\rm\sst M2}_{d}\equiv \CZ^{\rm \sst M2}_{\hat\si,\varpi}$ in the presence of $d$ codimension 4 surface
operators depends on the choice of a refined pants decomposition $\hat\si$, decorated with
certain additional discrete data collectively
denoted $\varpi$, see \cite{DGL}.  In \cite{AGGTV} it was
conjectured that the instanton partition functions $\CZ^{\rm\sst
  M2}_{\hat\si,\varpi}$ coincide with Liouville conformal blocks with $d$
additional degenerate fields inserted,
\begin{equation}\label{AGGTV}
\CZ^{\rm \sst M2}_{\hat\si,\varpi}(a,u,\tau;\ep_1,\ep_2)\,=\,\CZ^{\rm\sst L}_{\hat\si,\varpi}(\b,u,\tau;b)\,,
\end{equation}
given that the parameters are related as
\begin{equation}
\beta_e\,=\,\frac{Q}{2}+{\mathrm i}\frac{a_e}{\sqrt{\ep_1\ep_2}}\,,\qquad b^2\,=\,\frac{\ep_1}{\ep_2}\,.
\end{equation}
Further evidence for \rf{AGGTV} and some of its generalizations
were discussed in \cite{DGL,BGZa,BGZb,Gomis:2014eya}.


\bigskip

Now we are ready to bring together the results of the previous
sections to demonstrate the IR duality of two 4d gauge theories with
surface operators and to link it to the separation of variables in CFT
and Hitchin system.

\subsection{Relation to the Hitchin system and
to the separation of variables}    \label{Hit4}

It is shown in the  Appendix \ref{Classlim} that \rf{AGGTV}  implies that
\begin{equation}\label{epto0'}
\log\CZ^{\rm \sst M2}_{}(a,u,\tau;\ep_1,\ep_2)\,\sim\,-\frac{1}{\ep_1\ep_2}
\CF(a,\tau)-\frac{1}{\ep_1} \tilde \CW^{\rm \sst M2}_{}(a,u,\tau)\,,
\end{equation}
as already proposed in \cite{AGGTV}. The function
$\tilde \CW^{\rm \sst M2}_{}(a,u,\tau)$ is given as
\begin{equation}\label{Wfromv}
\tilde \CW^{\rm \sst M2}_{}(a,u,\tau)\,=\,-\sum_{k=1}^{h}\int^{u_k}v \,.
\end{equation}
We are now going to explain that there exist other sets of natural Darboux-coordinates $(u,v)$
for Hitchin moduli space allowing us to identify the function $\tilde \CW^{\rm \sst M2}_{}(a,u,\tau)$ defined
in \rf{Wfromv} as the generating function for the change of variables $(a,t)\leftrightarrow (u,v)$.

Recall from Section \ref{hit1} that the spectral cover construction
allows us to describe $\CM_{\rm \sst H}(C)$ as the space of pairs
$(\Sigma,L)$. The line bundle $L$ may be characterized by a divisor of
zeros of a particular section of $L$ representing a suitably
normalized eigenvector of the Higgs field $\vf \in
H^0(C,\on{End}({\mathcal E}) \otimes K_C)$ that we describe presently. Even
though this divisor is not unique, it's projection onto $C$ is
uniquely determined by the data of the rank two bundle ${\mathcal B}$
with a fixed determinant\footnote{As explained in Appendix \ref{cplxDarboux},
a natural possibility is to consider rank
  two bundles $\CB$  whose determinant is a fixed line bundle
  of degree $2g-2+n$. The moduli space of such bundles is isomorphic
  to the moduli space of $SL_2$-bundles on $C$.}  and the Higgs field
$\vf$.

Locally on $C$, we can trivialize the bundle ${\mathcal B}$ and choose
a local coordinate $z$. Then we can write $\vf$ as
$$
\vf = \bigg(\begin{matrix} a(z) & b(z) \\ c(z) & -a(z) \end{matrix} \bigg)dz.
$$
We have the following explicit formula for the  eigenvectors of $\vf$
$$
\Psi_\pm = \bigg(\begin{matrix} a(y)\pm v(y) \\
  c(z) \end{matrix}\bigg),\qquad v^2(y)=\frac{1}{2}{\rm tr}(\vf^2(y))\,.
$$
Note that for the matrix element $c(z) dz$ to be well-defined globally
on $C$ and independent of any choices, we need to represent ${\mathcal
  B}$ as an extension of two line bundles, see Appendix \ref{cplxDarboux}
  for more details.


If $c(z) \neq 0$, then $\Psi \neq 0$ for either branch of the square
root. If $c(z)=0$, then one of them vanishes. Now recall that the line
bundle $L$ on the double cover $\Sigma$ of $C$ is defined precisely as
the line bundle spanned by eigenvectors of $\vf$ (at a generic point
$p$ of $C$, $\varphi$ has two distinct eigenvalues, which correspond
to the two points, $p'$ and $p''$, of $\Sigma$ that project onto $p$,
and the fibers of $L$ over $p'$ and $p''$ are the corresponding
eigenvectors). Therefore, if we denote by ${\mathcal D}$ the divisor
of zeros of $c(z) dz$ on $C$, $\Psi$ gives rise to a non-zero section
of $L$ outside of the preimage of ${\mathcal D}$ in $\Sigma$.

Generically, ${\mathcal D}$ is multiplicity-free and hence may be
represented by a collection $u=(u_1,\dots,u_{d})$ of $d:={\rm deg}({\mathcal D})$
distinct points. The number number $d$ depends on the degrees of the line bundles used
to represent $\CB$ as an extension, in general. It may be larger than $3g-3+n$, the dimension
of ${\rm Bun}_G$. However, fixing the determinant of $\CB$ defines a collection of constraints
allowing us to determine $u_k$, $k=h+1,\dots,d$ in terms of the  coordinates $u_i$, $i=1,\dots,u_h$.

There are two distinct points, $u'_i$ and
$u''_i$, in $\Sigma$ over each $u_i \in C$. Then for each
$i=1,\ldots,h$, our section has a non-zero value at one of the points,
$u'_i$ or $u''_i$, and vanishes at another point. Thus, the divisor of
this section on $\Sigma$ is the sum of particular preimage of the
points $u_i, i=1,\ldots,h$, in $\Sigma$, one for each $i$. While there
is a finite ambiguity remaining for this divisor,\footnote{More
  precisely, we have $2^{3g-3+n}$ choices of the preimages $u'_i$ or
  $u''_i$ for each $i$, which agrees with the number of points in a
  generic Hitchin fiber corresponding to a fixed $SL_2$ bundle.} the
unordered collection $u=(u_1,\dots,u_{h})$ of points of $C$ is
well-defined (generically). And then for each $u_i$ we choose the
eigenvalue $v_k\in T_i^*C$, for which our section provides a non-zero
eigenvector. It is known that the collection
$(u,v)=((u_1,v_1),\dots,(u_{h},v_{h}))$ can be used to get to a system of
Darboux coordinates for $\CM_{\rm \sst H}(C)$ \cite{Hurt,GNR}, see
also \cite{Krich} for related results.

It was observed in \cite{GNR} that
the definition of the variables
$(u,v)$ outlined above can be seen as a generalization of the
method called  {\em separation of variables} in the literature on integrable models \cite{Skl}.
A familiar example is the so-called Gaudin-model
which can be identified with the Hitchin integrable system associated to surfaces $C$  of genus zero
with $n$  regular
singularities at distinct points $z_1,\ldots,z_n$.  The Higgs field can then be represented explicitly as
$$
\vf = \sum_{r=1}^n \frac{A_r}{y-z_i} dy, \qquad \sum_{r=1}^n A_r = 0,
$$
where
$$
A_r = \bigg(\,\begin{matrix} A_r^0 & A_r^+ \\ A_r^- &
  -A_r^0 \end{matrix} \, \bigg),
$$
and the separated variables are obtained as the zeros of the lower
left entry $A^-(y) dy$ of $\vf$:
\begin{subequations}\label{clSOV}
\begin{align}
&A^-(y)\,=\,u\frac{\prod_{k=1}^{n-3}(y-u_k)}{\prod_{r=1}^{n-1}(y-z_r)}\,,\\
&v_k = \sum_{r=1}^{n-1}\frac{A^0_r}{u_k-z_r}\,.
\end{align}
\end{subequations}

One may think of the separation of variables as a useful intermediate step in the construction
of the mapping from the original formulation of an integrable model to
the description as the Hitchin fibration in terms of action-angle
coordinates $(a,t)$. The remaining step from the separated variables
$(u,v)$ to the action-angle variables is then provided by the Abel
map. The function $\tilde \CW^{\rm \sst M2}_{}(a,u,\tau)$ is nothing
but the generating function for the change of Darboux coordinates
between $(u,v)$ and $(a,t)$. A few more details can be found in
Appendix \ref{Conn-to-Higgs}.

\subsection{IR duality of surface operators from
  the defects of codimension 2 and 4}

In this section we combine the ingredients of the brane analysis in
Section \ref{branecon} with our results on the twisted superpotentials
to show that the
4d gauge theories with the surface operators constructed from
codimension-2 and codimension-4 defects are equivalent in the IR.

Indeed, their vacuum structures are controlled by the twisted
superpotentials $\tilde \CW^{{\rm\sst M5}}(a,x,\tau)$ and $\tilde
\CW^{\rm\sst M2}(a,u,\tau)$, and we have found that they are related
by a change of variables (that is, a redefinition of fields).

Furthermore, when combined, the above arguments -- including the brane
creation upon the change of separation in the $x^7$ direction -- show
that two types of surface operators constructed from codimension-2 and
codimension-4 defects preserve the same supersymmetry subalgebra and
have the same twisted chiral rings.\footnote{Twisted chiral rings are
  Jacobi rings of the twisted chiral superpotential $\tilde \CW$ which
  has been our main subject of discussion in earlier sections.} This
is sufficient to establish their equivalence for the purposes of
instanton counting. In order to demonstrate the IR equivalence of the
full physical theories, we need to show the isomorphism between their
chiral rings (and not just the twisted chiral rings).  In general,
this is not guaranteed by the arguments we have used, but the good
news is that for simple types of surface operators, including the ones
considered here, the chiral rings are in fact trivial\footnote{In
  general, 2d $\CN=(2,2)$ theories may have non-trivial chiral and
  twisted chiral rings, see for example \cite{Lerche:1989uy}. However,
  if we start with a 2d theory without superpotential, then, as long
  as chiral superfields are all massive in the IR, integrating them
  out leads to a theory of twisted chiral superfields with a twisted
  superpotential, and so the chiral ring is indeed trivial.} and,
therefore, we do obtain the equivalence of the two full physical
theories.

As we already mentioned in the Introduction, this equivalence, or
duality, between the IR physics of 4d $\CN=2$ gauge theories with two
types of surface operators is conceptually similar to the Seiberg
duality of 4d $\CN=1$ gauge theories \cite{Seiberg:1994pq}. In fact,
it would not be surprising if there were a more direct connection
between the two phenomena since they both enjoy the same amount of
supersymmetry and in its brane realization, Seiberg's duality involves
the same kind of ``moves'' as the ones described in the previous
section.

\subsection{Turning on the Omega-deformation}\label{Liou-WZW}

The relation between $\tilde \CW^{{\rm\sst M5}}(a,x,\tau)$ and $\tilde
\CW^{\rm\sst M2}(a,u,\tau)$ has a rather nontrivial generalization in
the case of non-vanishing Omega-deformation that we will
describe in this subsection.
The fact that in 2d this a variant to the separation
of variables continues to hold for non-zero values of
$\epsilon_1$ and $\epsilon_2$ suggests that the two 4d $\CN=2$ gauge
theories remain IR equivalent even after Omega-deformation. The
possibility of such an equivalence certainly deserves further study.

When we quantize the Hitchin system, the separation of variables may
also be quantized.
In the genus zero case, in which the quantum
Hitchin system is known as the Gaudin model, this was first shown by
E. Sklyanin \cite{Skl}.
Note that the quantization of the classical Hitchin system
corresponds, from the 4d point of view, to ``turning on'' one of the
parameters of the Omega-deformation which is the case studied in \cite{Nekrasov:2009rc}.
It has been explained in Section 6 of
\cite{F:icmp} that one may interpret the separation of variables in
the Gaudin model, as well as more general quantum Hitchin systems, as
the equivalence of two constructions of the geometric Langlands
correspondence (Drinfeld's ``first construction'' and the
Beilinson--Drinfeld construction).

Feigin, Frenkel, and Stoyanovsky
have shown (see \cite{St}) that in genus zero the separation of variables of
the quantum Hitchin system maybe further deformed when we ``turn on''
both parameters of the Omega deformation. This result was subsequently generalized
to get relations between non-chiral correlation functions of the WZW-model and the
Liouville theory in genus 0 \cite{Ribault:2005wp}, and in higher genus \cite{Hikida:2007tq}.
It has furthermore been extended in \cite{Teschner:2010je} to larger classes of conformal
blocks.
{}From the 4d point of view,
this relation amounts to a rather non-trivial relation via an integral
transform (a kind of ``Fourier transform'') between the instanton
partition functions of the Omega-deformed 4d theories with surface
operators corresponding to the defects of codimensions 2 and 4.

The resulting relation has its roots in the quantum Drinfeld--Sokolov
reduction. We recall \cite{FF,FB} that locally it amounts to imposing
the constraint $J^-(z) = 1$ on one of the nilpotent currents of the
affine Kac--Moody algebra $\widehat{{\mathfrak s}{\mathfrak
    l}}_2$. The resulting chiral (or vertex) algebra is the Virasoro
algebra. Furthermore, if the level of $\widehat{{\mathfrak
    s}{\mathfrak l}}_2$ is
$$
k = -2 - \frac{1}{b^2},
$$
then the central charge of the Virasoro algebra is
$$
c = 1 + 6(b+b^{-1})^2.
$$

Globally, on a Riemann surface $C$, the constraint takes the form
$J^-(z) dz = \omega$, where $\omega$ is a one-form, if we consider the
trivial $SL_2$-bundle, or a section of a line bundle if we consider a
non-trivial $SL_2$-bundle that is an extension of two line sub-bundles
(the representation as an extension is necessary in order to specify
globally and unambiguously the current $J^-(z) dz$). Generically,
$\omega$ has simple zeros, which leads to the insertion at those
points of the degenerate fields $V_{-1/2b}$ of the Virasoro algebra in
the conformal blocks.

It is important to remember that classically the separated variables
$u_i$ are the zeros of a particular component of the Higgs field
$\vf$. But the Higgs fields correspond to the cotangent directions on
$\CM_{\rm \sst H}(C)$, parametrized by the $p$-variables. After
quantization, these variables are realized as the derivatives of the
coordinates along the moduli of $SL_2$-bundles (the $x$-variables), so
we cannot directly impose this vanishing condition. Therefore, in
order to define the separated variables $u$ in the quantum case, we
must first apply the Fourier transform making the $p$-variables into
functions rather than derivatives (this is already needed at the level
of the quantum Hitchin system, see \cite{F:icmp}). Since the Fourier
transform is an integral transform, our formulas below involve
integration. Indeed, the separation of variables linking the chiral
partition functions in the WZW-model and the Liouville model is an
integral transform.

In Appendix \ref{App-expl} it is shown that the relations described above can
be used to derive the following explicit integral transformation,
\begin{align}\label{SOVtrsf-M}
\check{\CZ}^{\rm\sst WZ}(x,z) &\,=\,N_J\int_{\ga} du_1\dots
du_{n-3}\;\CK^{\rm\sst SOV}(x,u)\, \check\CZ^{\rm\sst L}(u,z)\,,
\end{align}
where $\check{\CZ}^{\rm\sst WZ}$ and $\check{\CZ}^{\rm\sst L}$ are
obtained from ${\CZ}^{\rm\sst WZ}$ and ${\CZ}^{\rm\sst L}$ by taking
the limit $z_n\ra\infty$, and the kernel $\CK_{\rm SOV}(x,u)$ is
defined as
\begin{align}
&\CK^{\rm\sst SOV}(x,u):=
\left[\,{\sum_{r=1}^{n-1}x_r
\frac{\prod_{k=1}^{n-3}(z_r-u_k)}{\prod_{s\neq r}^{n-1}
(z_r-z_s)}}\right]^J\,
\prod_{k<l}^{n-3}(u_k-u_l)^{1+\frac{1}{2b^2}}
\prod_{r=1}^{n-1}\Bigg[\frac{\prod_{s\neq r}^{n-1}
  (z_r-z_s)}{\prod_{k=1}^{n-3}(z_r-u_k)}\Bigg]^{\al_r/b}\,;
\notag\end{align} $N_J$ is an $(x,z)$-independent normalization factor
that will not be needed in the following.  Note that the
$x$-dependence it entirely in the first factor on the right hand side
of \rf{KSOVdef}. Using \rf{epto0}, \rf{epto0'} and \rf{epto0''} it is
easy to see that the relation \rf{W-rel} follows from
\rf{SOVtrsf-M}. Formula \rf{SOVtrsf-M} is the relation
\eqref{SOV-1} discussed in the Introduction made explicit.

Thus, we see that the separation of variables in the most general case
(with both parameters of the Omega deformation being non-zero), viewed
as a relation between the chiral chiral partition functions in the
WZW-model and the Liouville model, provides the most satisfying
conceptual explanation of the IR duality of the 4d gauge theories with
surface operators of two kinds discussed in this paper.

\bigskip

\begin{center}

{\large {\bf APPENDICES}}

\end{center}

\appendix

\section{Surface operators and Nahm poles}
\label{Nahm}

Complex (co)adjoint orbits are ubiquitous in the study of both
half-BPS surface operators and boundary conditions.  This happens for
a good reason, and here we present a simple intuitive explanation of
this fact.  In short, it's due to the fact that both half-BPS surface
operators and boundary conditions are labeled by solutions to Nahm
equations.  Then, the celebrated work of Kronheimer \cite{Kronheimer}
relates the latter to complex coadjoint orbits.

\FIGURE{
\includegraphics[width=5.0in]{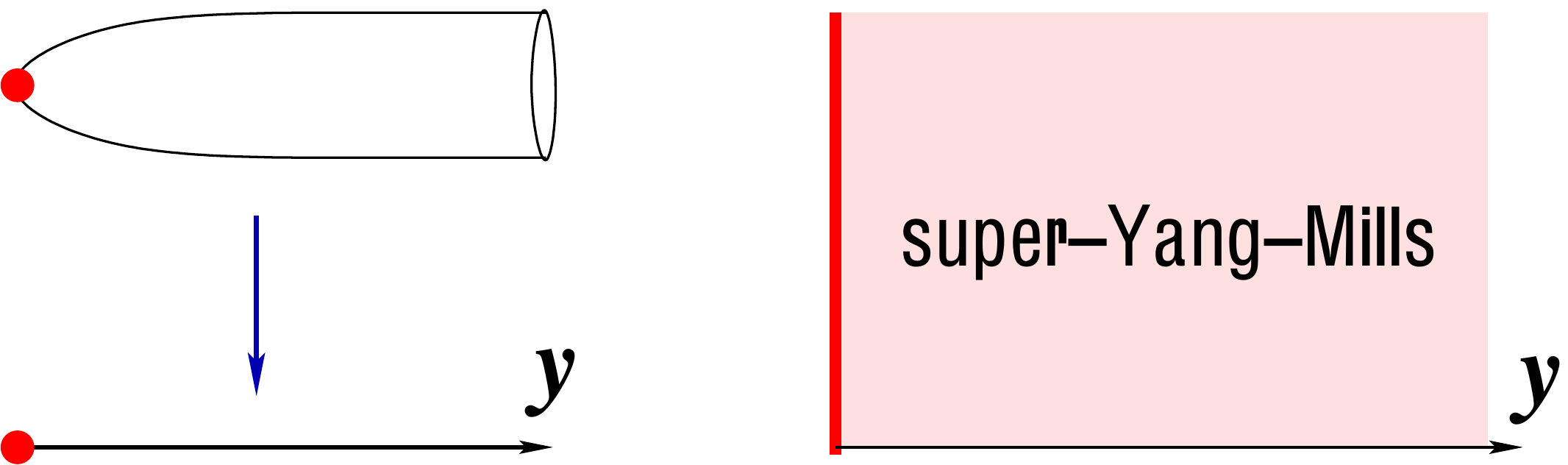}
  \caption{The six-dimensional $(2,0)$ theory with a codimension-2
    defect at the tip of the cigar reduces to 5d super-Yang-Mills
    theory with a non-trivial boundary condition.\label{cigarfig}}
}

Suppose that in our setup \eqref{M5M5} we take $C={\bf S}^1_C \times
\mathbb{R}$ and $M_4 = D \times D' \cong \mathbb{R}^4$, where $D'
\cong \mathbb{R}^2$ is the ``cigar.''  In other words, $D'$ is a
circle fibration over the half-line, $\mathbb{R}_+ = \{ y \ge 0 \}$,
with a singular fiber at $y=0$ so that asymptotically (for $y \to +
\infty$) $D$ looks like a cylinder, see Figure~\ref{cigarfig}.  Then,
the six-dimensional $(2,0)$ theory on $M_4 \times C$ with a
codimension-2 defect on $D \times C$ can be reduced to
five-dimensional super-Yang-Mills theory in two different ways.
First, if we reduce on a circle ${\bf S}^1_C$, we obtain a 5d
super-Yang-Mills on $M_4 \times \mathbb{R} \cong \mathbb{R}^5$ with a
surface operator supported on $D \times \mathbb{R} \cong
\mathbb{R}^3$.  If we denote by $r = e^{-y}$ the radial coordinate in
the plane transverse to the surface operator, then the supersymmetry
equations take the form of Nahm's equations:
\begin{align}\label{Nahmeqs}
 \frac{da}{dy} \,=\, [b,c] \,, \qquad
 \frac{db}{dy} \,=\, [c,a] \,, \qquad
 \frac{dc}{dy} \,=\, [a,b] \,,
\end{align}
where we used the following ansatz for the gauge field and for the
Higgs field:
\begin{align}
& A \; = \; a(r) d \theta \,, \qquad
\phi \; = \; b(r) \frac{dr}{r} + c(r) d \theta \,. \notag
\end{align}
On the other hand, if we first reduce on the circle fiber ${\bf
  S}^1_F$ of the cigar geometry $D'$, we obtain a 5d super-Yang-Mills
on $\mathbb{R}_+ \times D \times C$ with a non-trivial boundary
conditions at $y=0$ determined by the codimension-2 defect of the
six-dimensional theory.  Note, these boundary conditions are also
associated with solutions to Nahm's equations \eqref{Nahmeqs} for the
Higgs field $\vec \phi = (a,b,c)$.  Further dimensional reductions of
these two systems yield many half-BPS boundary conditions and surface
operators in lower-dimensional theories, all labeled by solutions to Nahm's equations.

\FIGURE{
\includegraphics[width=2.5in]{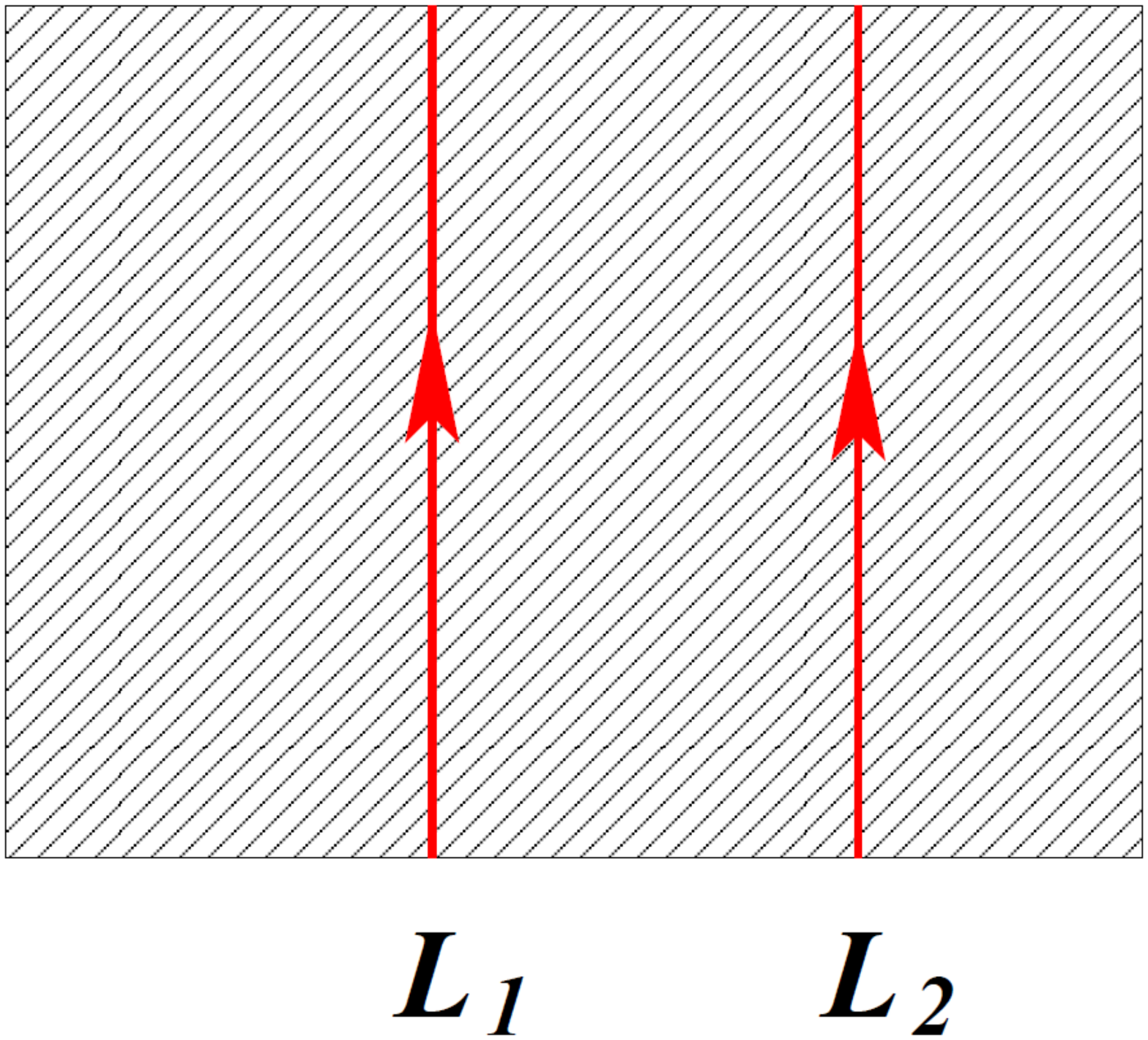}
\caption{In the presence of surface operator and/or Omega-background
  line operators do not commute.\\}
\label{linefig}
}

Among other things, this duality implies that similar physical and mathematical structures can be found on
surface operators as well as in the study of boundaries and interfaces.
A prominent example of such structure is the algebra of parameter walls and interfaces,
{\it i.e.} Janus-like solitons realized by monodromies in the space of parameters.
(In the case of surface operators, such monodromy interfaces are simply line operators,
which in general form non-commutative algebra if they can't move off the surface operator, as illustrated in Figure \ref{linefig}.)

This description of walls, lines and interfaces as monodromies in the parameter space provides
a simple and intuitive way of understanding their non-commutative structure and commutation relations;
it is captured by the fundamental group of the parameter space \cite{Gukov:2006jk}:
\be
\pi_1 (\{ \text{parameters} \})
\ee
For instance, in the case of $C = T^2$ one finds $\pi_1 \left( (\mathbb{T}_{\C} / S_N)^{\text{reg}} \right)$,
which is precisely the braid group (in the case, of type $A_{N-1}$).
It is generated by parameter walls / interfaces $L_i$ that obey the standard braid group relations:
\be
L_i \star L_{i+1} \star L_i \; = \; L_{i+1} \star L_{i} \star L_{i+1}
\ee
{}From 2d and 3d perspectives, these systems are often described by sigma-models based
on flag target manifolds (or their cotangent bundles) where the lines/walls $L_i$
are represented by twist functors; see \cite{Gukov:2007ck,Gukov:2014gja} for further details and
many concrete examples of braid group actions on boundary conditions.
The case of the parameter space \eqref{SymC} is qualitatively similar.

\section{Twisting of Kac-Moody conformal blocks}\label{twistapp}

This appendix collects some relevant mathematical background concerning the dependence of
Kac-Moody conformal blocks on the choice of a holomorphic bundle on $C$.

\subsection{Twisted conformal blocks}\label{twistbasics}

A generalization of the defining invariance condition allows us to define a generalized notion of conformal
blocks depending on the choice of a holomorphic $G$-bundle $\CB$ on $C$.
One may modify the defining invariance condition \rf{Ward} by replacing the elements of the Lie algebra
$\gg_{\out}$ by a section of
\begin{equation}
\gg_{\out}^{\CB}:=\Ga(C,\gg_\CB^{})\,,\qquad \gg_\CB^{}:=\CB\underset{G}\times\gg\,.
\end{equation}
Describing $\CB$ in terms of a cover $\{\CU_\imath;\imath\in\CI\}$ of $C$ allows us to describe
$\CB$  in terms of the
$G$-valued transition functions $h_{\imath\jmath}(z)$ defined on the intersections
$\CU_{\imath\jmath}=\CU_{\imath}\cap\CU_{\jmath}$. The sections of $\gg_{\out}^{\CB}$ are
represented by  families of $\gg$-valued functions
$\eta_{\imath}$  in $\CU_\imath$, with $\eta_{\imath}$  and $\eta_{\jmath}$  related
on the intersections $\CU_{\imath\jmath}$ by conjugation with
$h_{\imath\jmath}(z)$. In this way one  defines $\CB$-twisted conformal blocks
$\varphi_{\CB}^{}$ depending on the choice of a $G$-bundle $\CB$.

More concrete ways of describing the twisting of conformal blocks are obtained by
choosing convenient covers $\{\CU_\imath;\imath\in\CI\}$. One convenient choice is the following:
Let us choose discs $\mathbb{D}_{k}$ around the points $z_k$, $k=1,\dots,n$
such that $\mathcal{U}_{\rm out}:=
C\setminus \{z_1,\dots,z_n\}$ and $\mathcal{U}_{\rm in}=\bigcup_{k=1}^n\mathbb{D}_k$ form a cover of $C$.
It is known that for $G=SL(2)$
$G$-bundles $\CB$ can always be trivialized in $\mathcal{U}_{\rm out}$ and $\mathcal{U}_{\rm in}$.
An arbitrary  $G$-bundle $\CB$
can then be represented by the $G$-valued transition functions $h_k(t_k)$ defined in
the annular regions $\mathbb{A}_k:=\mathcal{U}_{\rm out}\cap
\mathbb{D}_{k}$ modulo changes of trivialization in $\mathcal{U}_{\rm in}$ and in
$\mathcal{U}_{\rm out}$, respectively.

Introducing the dependence on the choice of $\CB$ in the way described above
makes it easy to see that infinitesimal variations $\de$ of $\CB$ can be represented by elements of
$\bigoplus_{i=1}^n \gg \otimes \C(\!(t_i)\!)$.
Choosing a lift $\mathsf{X}_\de$ to the diagonal central
extension of $\bigoplus_{i=1}^n \gg \otimes \C(\!(t_i)\!)$ allows us to define a (projective)
action of $T{\rm Bun}_{G}\big|_{\CB}$
on ${\rm CB}_{\gg}(\CR_{1},\ldots,\CR_n)$.
This means that a differential operator $\de$ representing an element $T{\rm Bun}_{G}\big|_{\CB}$
can be represented on the conformal blocks in terms of the action
of $\eta_\de$ on $\bigotimes_{r=1}^n\CR_r$, schematically
\begin{equation}\label{currentWard}
\de\vf(e_{[n]})\,=\,\vf\big(\,\eta_\de^{} \,e_{[n]}\big)\,,\qquad e_{[n]}:=e_1\otimes\dots\otimes e_n\,.
\end{equation}
This action describes the response of a
conformal block $\vf_{\CB}$ with respect to an infinitesimal
variation of $\CB$.

\subsection{Genus zero case}

In the case of genus $0$ it suffices to choose the
transition functions $h_k(t_k)$ in
the annular regions $\mathbb{A}_k$ around the points $z_k$ to be the constant nilpotent matrices
$h_k(t_k)=\left(\begin{smallmatrix} 1 & x_k \\ 0 & 1
\end{smallmatrix}\right)$. The collection of parameters $x=(x_1,\dots,x_n)$ can be used to
represent the dependence on the choice of $\CB$ in this case.
The action of $T{\rm Bun}_{G}\big|_{\CB}$
on spaces of conformal blocks defined via  \rf{currentWard} may then be
represented more explicitly in terms of the differential operators
${\CJ}^a_r$ defined as \begin{equation}\label{CJdef}
{\CJ}^-_r=\pa_{x_r} ,\quad
{\CJ}^0_r=x_r\pa_{x_r}-j_r,\quad
{\CJ}^+_r=-x^2_r\pa_{x_r}+2j_rx_r\,.
\end{equation}
The Casimir operator is represented as multiplication by $j_r(j_r+1)$.

The parametrization in terms of $n$ variables $x=(x_1,\dots,x_n)$ is of course
redundant. The conformal Ward-identities \rf{Ward} include the invariance under global
$\fsl_2$-transformations, allowing us to eliminate three out of the $n$ variables
$x_1,\dots,x_n$ in the usual way.

The operators $\SH_r$ appearing in the Knizhnik-Zamolodchikov equations \rf{KZ}
are then given  by the formulae
\begin{equation}\label{Hdef}
\mathsf{H}_r \equiv \sum_{s\neq r}\frac{{\CJ}_{rs}}{{z}_r-{z}_s}\,,
\end{equation}
where the differential operator ${\CJ}_{rs}$ is defined as
\begin{equation}\label{kzmu}
{\CJ}_{rs}:=\eta_{aa'}{\CJ_r^a\CJ_s^{a'}} :=
{\CJ}^0_r {\CJ}^0_s + \frac{1}{2} ({\CJ}^+_r {\CJ}^-_s +{\CJ}^-_r {\CJ}^+_s)\,.
\end{equation}
The operators $\SH_r$ commute, and may therefore be used as
Hamiltonians for generalizations of the Gaudin models associated to more
general representations of $SL(2,\BC)$.

\subsection{Higher genus cases}

Instead of the covers considered in Subsection \ref{twistbasics} above
one may use alternatively use covers defined using the gluing construction.
One thereby gets a cover
$\{\CU_\imath;\imath\in\CI\}$ with intersections represented by
annuli $\mathbb{A}_e$ between pairs of pants or connecting two legs of the same pair of pants.
Choosing constant
diagonal transition functions $\big(\begin{smallmatrix} x_e & 0 \\ 0 & x_e^{-1} \end{smallmatrix}\big)$ in
the annuli $\mathbb{A}_e$ gives us a collection of local coordinates $x_e$, $e=1,\dots,3g-3+n$
 for ${\rm Bun}_{G}$, $G=SL(2)$.   The resulting parameters $x$ for ${\rm Bun}_G$ are
 easily identified with
the parameters $x$ introduced in the gluing construction of conformal blocks
via \rf{gluing}
provided we choose  $\CK(\tau,x)$ to be $e^{2\pi i \tau L_0}x^{J_0^0}$.

In order to have a globally well-defined current $J^-$ on $C$ one needs to represent
$\CB$ as an extension. Taking
\begin{equation}
0\longrightarrow \CO \longrightarrow \CB\longrightarrow  \CL \longrightarrow 0\,,
\end{equation}
appears to be particularly natural.
This allows us to represent $J^-$ as a section of $\CL\ot K_C$. As explained
in  Appendix \ref{Hitchbackgr} it is natural in our case to consider fixed  line bundles $\CL$ of degree
$d'$. Let us represent $\CL$ as $\CO(\CD')$, with divisor $\CD'$ being represented by the
points $y_1,\dots y_{d'}$. The bundle $\CB$ may
be described by using a cover $\{\CU_\imath;\imath\in\CI\}$ for $C$ containing small discs $\BD_k'$
around $y_k$, $k=1,\dots,d'$, with transition functions
\begin{equation}\label{Hecke}
h_k'=
\bigg(\begin{matrix} 1 & 0 \\ 0 & t_k
\end{matrix}\bigg)
\bigg(\begin{matrix} 1 & x_k \\ 0 & 1
\end{matrix}\bigg)\,,
\end{equation}
on the annuli ${\mathbb A}_k'=\BD_k'\setminus\{y_k\}$, where $t_k$ is a coordinate on $\BD_k'$ vanishing at $y_k$.
Sections of $\CB$ may alternatively be
represented locally by functions that are regular outside of $\{ y_k,k=1,\dots,d'\}$
and may have poles with residue in a fixed line $\ell_k$ at $y_k$, $k=1,\dots,d'$.
Using the transition functions \rf{Hecke} determines the lines $\ell_k$ in terms of the parameters
$x_k$. Modifications of $\CB$ that increase the degree $d'$ of $\CL$ are called Hecke modifications.

Using covers defined with the help of the gluing construction it appears to be natural to take $d'=2g-2$.
In this case one may assume that
there is exactly one $y_k$ contained in each pair of pants. Kac-Moody conformal blocks  associated
to each pairs of pants appearing in the pants decomposition
of a closed Riemann surface can then be defined using conformal blocks on $C_{0,4}$,
with one insertion being the degenerate representation of the Kac-Moody algebra
$\CR_{k/2}$ representing the Hecke modifications within conformal field theory \cite{Teschner:2010je}.
If the Riemann surface has punctures, one may use conformal blocks on $C_{0,3}$ without
extra insertion of $\CR_{k/2}$ for the pairs of pants containing the punctures.

It is worth remarking that $d'=2g-2$ is exactly the case where the
current $J^-$, being a section of $K_C\otimes \CL$, has $4g-4$ zeros
$u_i$, as required by the identification of the points $u_i$ with the
end-points of the M2-branes created from the M5$'$-branes.

\section{Holomorphic pictures for the Hitchin moduli spaces}\label{Hitchbackgr}

The Hitchin space $\CM_{\rm\sst H}(C)$
was introduced in the main text as the space of pairs $(\CB,\vf)$. Interpreting the
Higgs fields $\vf\in H^0(C,{\rm
  End}(\CE)\otimes K_C)$ as representatives of cotangent vectors to ${\rm Bun}_{G}$,
  one may  identify
$\CM_{\rm\sst H}(C)$ with $T^*{\rm Bun}_{G}$, the cotangent bundle of the moduli space of
holomorphic $G$-bundles on $C$. This description equips $\CM_{\rm\sst H}(C)$
with natural complex and symplectic structures, leading to the definition of local
sets of Darboux coordinates $(x,p)$ parametrizing the choices of $G$-bundles via
coordinates $x$, and the choices of Higgs fields $\vf$ in terms of holomorphic coordinates $p$.

In order to exhibit the relation with conformal field theory we
will find it, following \cite{BF,Teschner:2010je},
useful to consider a family of other models for  $\CM_{\rm\sst H}(C)$.
We will consider
moduli spaces $\CM_{\rm\sst H}^{\ep}(C)$  of pairs $(\CB,\nabla'_\ep)$ consisting of
holomorphic  bundles $\CB$ with holomorphic $\ep$-connections $\nabla'_\ep$.  An $\ep$-connection
is locally represented by a differential operator $\nabla'_\ep=(\ep\pa_y+A(y))dy$
transforming as $\tilde\nabla'_\ep=g^{-1}\cdot\nabla'_\ep\cdot g$ under gauge-transformations.
Consideration of $\CM_{\rm\sst H}^\ep(C)$ will represent a useful intermediate step
which helps clarifying the link between conformal field theory and the
Hitchin system.
Noting that any two $\ep$-connections $\nabla'_\ep$  and $\tilde\nabla'_\ep$ differ by an element
of $H^0(C,{\rm End}(\CE)\otimes K_C)$ one sees that $\CM_{\rm\sst H}^\ep(C)$ can be regarded
as a twisted cotangent bundle $T^*_\ep{\rm Bun}_G$. Picking a reference connection
$\nabla'_{\ep,0}$, one may represent a generic connection as $\nabla'_\ep=\nabla'_{\ep,0}+\vf$.

To avoid confusion let us stress that the
resulting isomorphism $\CM_{\rm\sst H}^\ep(C)\simeq
T^*{\rm Bun}_{G}$
is not canonical, being dependent on the choice of $\nabla'_{\ep,0}$. Instead we could
use the known results of Hitchin, Donaldson, Corlette and Simpson \cite{Hi87,Don87,Cor,Sim88,Sim92}
relating pairs $(\CB,\vf)$ to flat connections on $C$ to identify
the moduli spaces $\CM_{\rm\sst H}(C)$ and $\CM_{\rm\sst H}^{\ep}(C)$.
The description of $\CM_{\rm\sst H}^\ep(C)$  as twisted cotangent bundle yields natural
complex and symplectic structures which are inequivalent for different values of $\ep$.
This can be used to describe the hyperk\"ahler structure on $\CM_{\rm\sst H}(C)$,
with $\ep$ being the hyperk\"ahler parameter \cite{Sim97}.

However, in order to discuss the relation with conformal field theory we find it useful
to adopt a different point of view. The definition of conformal blocks depends on the choice of
a $G$-bundle $\CB$, which may be parametrized by variables $x$ in a way that
does not depend on $\ep_1$ and $\ep_2$.  The gluing construction yields natural choices for
the reference connection  $\nabla'_{\ep,0}$, e.g. the trivial one. All dependence on the parameter
$\ep$ is thereby shifted into the relations between different charts $\CU_\imath$
on $\CM_{\rm\sst H}^\ep(C)$ parametrized in terms of local coordinates $(x_\imath,p_\imath)$
in a way that does not explicitly depend on $\ep$.

One may formally identify $\vf\in H^0(C,{\rm End}(\CE)\otimes K_C)$ as an $\ep$-connection for $\ep=0$.
We therefore expect that the Darboux coordinates $(x_\ep,p_\ep)$ turn into the
Darboux coordinates $(x,p)$ discussed in the main text when $\ep\ra 0$. This will be further
discussed below, after having discussed possible choices of Darboux coordinates more
concretely.

\subsection{Three models for Hitchin moduli space}
There are three models
for $\CM_{\rm\sst H}^{\ep}(C)$ of
interest for us:
\begin{itemize}
\item[(A)] As space of representations of the fundamental group
\begin{equation}
{\rm Hom}(\pi_1(C),{\rm SL}(2,\BC))/{\rm SL}(2,\BC)\,.
\end{equation}
\item[(B)] As space of bundles with connections $(\CE,\nabla'_\ep)$,
\begin{equation}\label{holoconn}
\nabla' _\ep= (\ep\pa_y+A(y))\,dy\,,\quad
A(y) = \bigg(\,\begin{matrix} A^0(y) & A^{+}(y) \\ A^-(y) & -A^0(y)
\end{matrix}\,\bigg)\,.
\end{equation}
Having $n$ punctures $z_1,\dots,z_n$ means that $A(y)$ is allowed to have regular
singularities at $y=z_r$ of the form
\begin{equation}
A(y)\,=\,\frac{A_r}{y-z_r}+\text{regular}\,.
\end{equation}
\item[(B')] As space of opers $\ep^2\pa_y^2+t(y)$, where $t(y)$ has $n$ regular singularities
at $y=z_r$,
\begin{equation}\label{t(y)g=0}
t(y)=\frac{\de_r}{(y-z_r)^2}-\frac{H_r}{y-z_r}
+\text{regular}\,,
\end{equation}
and $d$ apparent singularities at $y=u_k$,
\begin{equation}
t(y)=-\frac{3\ep^2}{4(y-u_k)^2}+
\frac{\ep v_k}{y-u_k}
+\text{regular}\,,
\end{equation}
Having an apparent singularity at $y=u_k$ means that the monodromy around $u_k$
is trivial in ${\rm PSL}(2,\BC)$. This is known \cite[Section 3.9]{F:icmp} to be equivalent to the fact that
the residues $H_r$, $r=1,\dots,n$ are constrained
by the linear equations
\begin{subequations}\label{Hconstr}
\begin{align}
& v_k^2+t_{k,2}\,=\,0\,,\quad k=1,\dots,l\,,\quad
t(y)=\sum_{l=0}t_{k,l}(y-u_k)^{l-2}\,.\label{cldec}\end{align}
If $g=0$, the parameters $H_s$, $s=1,\dots,n$ are furthermore constrained by
\begin{equation}
\sum_{r=1}^n z_r^a(z_rH_r+(a+1)\de_r)=0\,,\qquad a=-1,0,1\,,
\label{projinv}
\end{equation}
\end{subequations}
ensuring regularity of $t(y)$ at infinity.
\end{itemize}
Models (B) and (B') are related by singular gauge
transformations which transform $A(y)$ to the
form
\begin{equation}\label{operconn}\tilde{A}(y)=
\bigg(\,\begin{matrix} 0 & -t(y) \\ 1 & 0
\end{matrix}\,\bigg)\,.
\end{equation}
In order to describe the relation between (B) and (B') more concretely let us,
without loss of generality, assume that
elements of ${\rm Bun}_G$ are represented as extensions
\begin{equation}\label{EXT}
0\longrightarrow \CL'\longrightarrow \CB\longrightarrow \CL''\longrightarrow 0\,.
\end{equation}
Describing the bundles $\CB$ by means of a covering $\CU_\imath$ of $C$ and
transition functions $\CB_{\imath\jmath}$ between patches $\CU_{\imath}$ and
$\CU_{\jmath}$, one may assume that all $\CE_{\imath\jmath}$ are upper triangular,
\begin{equation}\label{transfct}
\CB_{\imath\jmath}\,=\,\bigg(\begin{matrix} \CL_{\imath\jmath}' & 0\\
0 & \CL_{\imath\jmath}''\end{matrix}\bigg)\bigg(\begin{matrix} 1 & \CE_{\imath\jmath}\\
0 & 1\end{matrix}\bigg)\,.
\end{equation}
This implies that the lower left matrix element $A^-(y)$ of the $\ep$-connection
$\ep\pa_y+A(y)$  is a section of the line bundle $(\CL')^{-1}\otimes\CL''\otimes K_C$,
with $K_C$ being the canonical line bundle.
The gauge transformation which transforms $A(y)$ to the form \rf{operconn} will be singular
at the zeros $u_k$ of $A^-(y)$,
leading to the appearance of the apparent singularities $u_k$ in
\rf{t(y)g=0}.

\subsection{Complex-structure dependent Darboux coordinates}\label{cplxDarboux}

Let us briefly discuss possible ways to introduce Darboux coordinates $(x,p)$ for $\CM_{\rm\sst H}^\ep(C)$,
and how the passage from $\ep$-connections to opers defines a change of Darboux coordinates
from $(x,p)$ to $(u,v)$.

\subsubsection*{Genus zero}
In the cases of genus $g=0$ we may parametrize the matrices
$A_r$ in \rf{holoconn} as
\begin{equation}\label{Mfactor}
  A_r\equiv\bigg(\,\begin{matrix} A_r^0 & A_r^+ \\ A_r^- &
    -A_r^0 \end{matrix}\,\bigg)
  \equiv\bigg(\begin{matrix} 1 & -x_r \\ 0 & 1
\end{matrix}\bigg)\bigg(\begin{matrix} l_r & 0 \\ p_r & -l_r
\end{matrix}\bigg)\bigg(\begin{matrix} 1 & x_r \\ 0 & 1
\end{matrix}\bigg)\,,
\end{equation}
assuming that $(x_r,p_r)$ are a set of Darboux coordinates
with $\{p_r,x_s\}=\de_{r,s}$. Let $\CP_n$ be the phase space whose
algebra of functions is generated by functions of $(x_r,p_r)$,
$r=1,\dots,n$. The space $\CM_{\rm flat}(C_{0,n})$
can be described as the symplectic reduction of $\CP_n$
w.r.t. the global $\fsl_2$-constraints
\begin{equation}\label{globalsl2}
\sum_{r=1}^n A_r^a=0\,,
\end{equation}
for $a=-,0,+$, or, more conveniently, as the
symplectic reduction of $\CP_{n-1}$
w.r.t. the constraints \rf{globalsl2} for $a=-,0$
combined with sending $z_n\ra\infty$.
We will use the latter
description.

The change of $(x,p)\leftrightarrow (u,v)$ induced by the relation
between models (B) and (B') is explicitly described by the formulas
(note that the same formulas \eqref{clSOV} appear in the limit
$\epsilon \to 0$):
\begin{subequations}\label{cl-SOV}
\begin{align}
&A^-(y)\,=\,u\frac{\prod_{k=1}^{n-3}(y-u_k)}{\prod_{r=1}^{n-1}(y-z_r)}\,,\\
&v_k:=\,A^0(u_k)\,,\qquad A^0(y)\,=\,\sum_{r=1}^{n-1}\frac{A^0_r}{y-z_r}\,.
\end{align}
\end{subequations}
The resulting change of variables $(x,p)\leftrightarrow(u,v)$ is known
to be a change of Darboux coordinates.  It is in fact the classical
version of the separation of variables transformation for the
Schlesinger system \cite{DuMa}.  In order to see this, let us consider
in the model (B') the case $l=n-3$. In this case the equations
\rf{Hconstr} determine the $H_r$ as functions of the parameters
$(u,v)$, $u=(u_1,\dots,u_l)$, $v=(v_1,\dots,v_l)$.  The solutions
$H_r(u,v;z)$ to the constraints \rf{Hconstr} are the Hamiltonians of
the Garnier system.  The flows generated by the Hamiltonians
$H_r(u,v;z)$ preserve the monodromy of the oper $\ep^2\pa_y^2+t(y)$.

In the model (B) one may consider
the Schlesinger Hamiltonians defined as
\begin{equation}
H_r(x,p;z):=\,\sum_{s\neq r}\eta_{ab}\frac{A_r^aA_s^b}{z_r-z_s}\,;
\end{equation}
It is well-known that the non-autonomous Hamiltonian flows generated by the $H_r$
preserve the monodromy of the connection $\ep\pa_y+A(y)$.
The change of variables defined via \rf{cl-SOV}
relates the Hamiltonians $H_r(x,p;z)$ to the Hamiltonians $H_r(u,v;z)$
of the Garnier system.

\subsubsection*{Higher genus}
Considering the cases of higher genus one may introduce Darboux coordinates associated
to the model (B) as follows. To simplify the discussion slightly let us consider closed Riemann surfaces,
$n=0$.
Representing the bundles $\CB$ as extensions \rf{EXT}, there are two places where the moduli
may hide, in general: They may be hidden in the choice of the line bundles $\CL'$, $\CL''$,
as well as in the extension classes $\CE\in H^1(\CL^{-1})$, in terms of transition functions represented by the
$\CE_{\imath\jmath}$
in \rf{transfct}. A particularly simple case is found by choosing
$\CL'=\CO$ and $\CL''\equiv \CL$ in \rf{EXT}, with $\CL$ being a
fixed line bundle of degree $2g-2$.
Fixing $\CL$ is equivalent to fixing the determinant of $\CB$.

The dimension of the
space of extension classes is then ${\rm dim}( H^1(\CL^{-1}))=g-1+{\rm deg}(\CL)=3g-3$.
The moduli of ${\rm Bun}_G$ can therefore be parametrized by the choices of extension classes.
Coordinates $x=(x_1,\dots,x_{3g-3})$ on $H^1(\CL^{-1})$ give coordinates for ${\rm Bun}_G$.

Serre duality implies that the dual of $H^1(\CL^{-1})$ is the space $H^0(\CL\ot K_C)$.
Recall that the lower left matrix element $A^-(y)$ of an $\ep$-connection
$\ep\pa_y+A(y)$  is a section of the line bundle $\CL\otimes K_C$. Finding
coordinates for $H^0(\CL\ot K_C)$ that are dual to the coordinates $x$
on $H^1(\CL^{-1})$ with respect to the pairing provided by
Serre duality will therefore give us coordinates $p=(p_1,\dots,p_{3g-3})$ that are
canonically conjugate to the coordinates $x$ on ${\rm Bun}_G$.

\subsection{Complex-structure independent
Darboux coordinates}\label{NRScoords}

Representing elements of  $\CM_{\rm flat}(C)$ in terms
of the model (A) mentioned above allows one to introduce useful
Darboux coordinates which do not depend on a choice of complex structure
of $C$ as opposed to the coordinates $(u,v)$ and $(x,p)$
introduced before.
A convenient description was given in
\cite{Nekrasov:2011bc} and references therein.

\FIGURE{\includegraphics[width=0.35\textwidth]{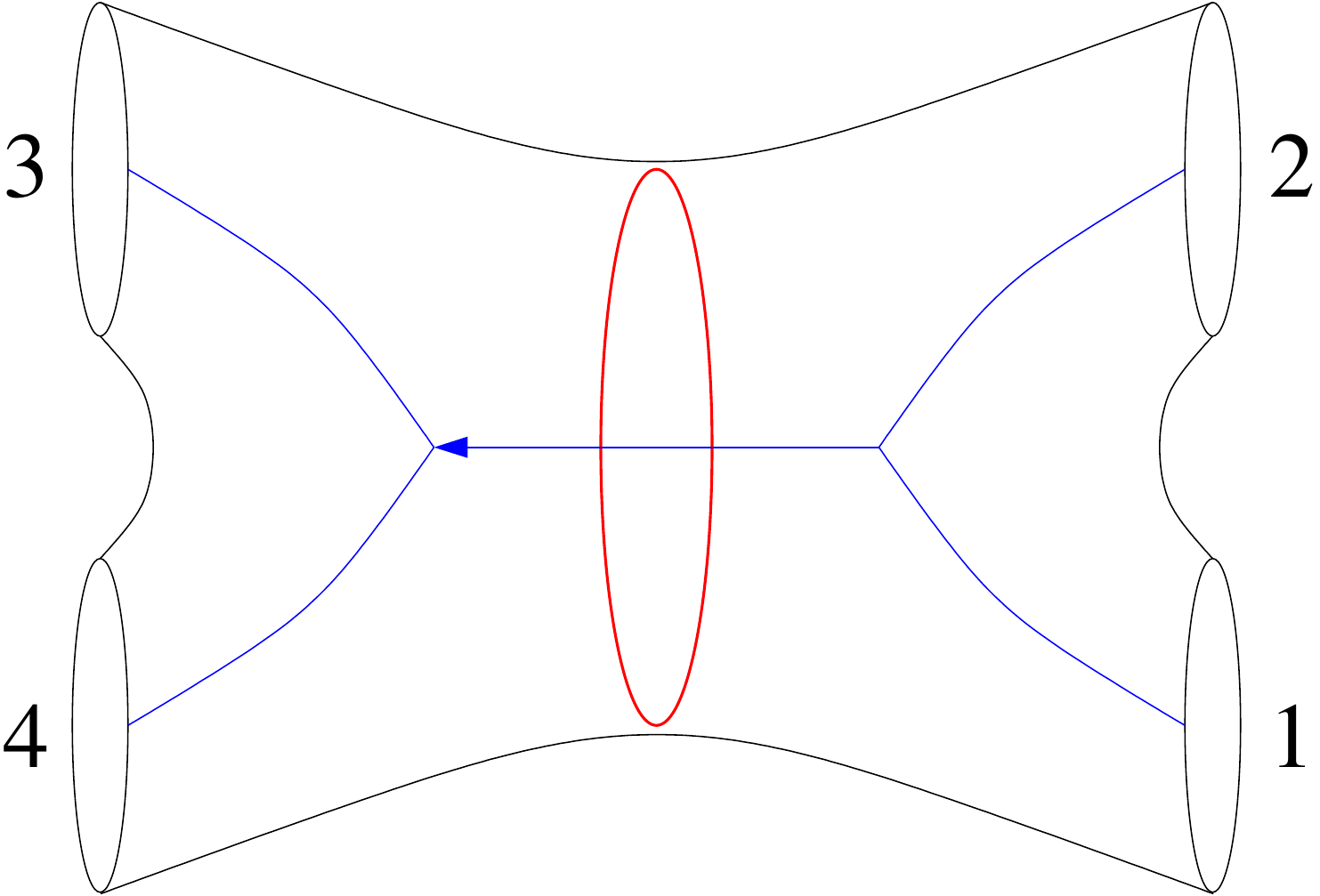}
\caption{Pants decomposition of four-holed sphere with a numbering of
  boundary components.}
\label{fmove}
}

Let us use the set-up from Section \ref{classS1}. A trivalent graph
$\si$ on $C$
determines a pants decomposition defined by cutting along
the simple closed curves $\ga_e$ which intersect the edge $e$ of $\si$
exactly once. For each (oriented) edge $e$ we shall denote
$\ga_{e,s}\equiv\ga_e$, $\ga_{e,t}$ and $\ga_{e,u}$
the simple closed curves which encircle the pairs of
boundary components $(\ga_{e,1},\ga_{e,2})$,
$(\ga_{e,2},\ga_{e,3})$ and $(\ga_{e,1},\ga_{e,3})$,
respectively, with labeling of boundary components introduced via Figure \ref{fmove}.
Let $L_{e,i}:={\rm tr}(\rho(\ga_{e,i}))$ for $i\in\{s,t,u,1,2,3,4\}$.
One may represent $L_{e,s}$, $L_{e,t}$ and $L_{e,u}$
in terms of Darboux coordinates $a_e$ and $k_e$ which have
Poisson bracket
\begin{equation}
\{\,a_e\,,\,k_{e'}\,\}\,=\,\frac{\ep^2}{(2\pi)^{2}}\de_{e,e'}\,.
\end{equation}
The expressions
are
\begin{subequations}\label{FN-FK}
\begin{align}
& L_{e,s}\,=\,2\cosh(2\pi a_e/\ep)\,,\\
& L_{e,t}\big((L_{e,s})^2-4\big)\,=\,
2(L_{e,2}L_{e,3}+L_{e,1}L_{e,4})+L_{e,s}(L_{e,1}L_{e,3}+L_{e,2}L_{e,4}) \label{cl-'t Hooft}\\
& \hspace{3cm}
+2\cosh(2\pi k_e/\ep)
\sqrt{c_{12}(L_{e,s})c_{34}(L_{e,s})}\,,
\notag\\
& L_{e,u}\big((L_{e,s})^2-4\big)\,=\,2(L_{e,1}L_{e,3}+L_{e,2}L_{e,4})+
L_{e,s}(L_{e,2}L_{e,3}+L_{e,1}L_{e,4}) \label{cl-dyonic}\\
& \hspace{3cm}
+2\cosh(\pi(2k_e-a_e)/\ep)
\sqrt{c_{12}(L_{e,s})c_{34}(L_{e,s})}\,,
\notag
\end{align}
\end{subequations}
where  $c_{ij}(L_s)$ is defined as
\begin{align}\label{cijdef}
c_{ij}(L_s) & \,=\,L_s^2+L_i^2+L_j^2+L_sL_iL_j-4\,.
\end{align}
Restricting these Darboux coordinates to the Teichm\"uller component
we recover the Fenchel-Nielsen length-twist coordinates
well-known in hyperbolic geometry.

\subsection{Limit $\ep\ra 0$: Recovering the Higgs pairs}\label{Conn-to-Higgs}

We now want to send $\ep\ra 0$. One may note that the equation $(\ep\pa_y+A(y))\psi(y;x,z)$ can
in the limit $\ep$ be solved to leading order in $\ep$ by an ansatz of the form
\begin{equation}\label{WKB-horizontal}
\psi(y;x,z)=
e^{-\frac{1}{\ep}\int^{y}du\,v(u)}\chi(y;x,z)\,,
\end{equation}
where $\chi(y;x,z)$ is an eigenvector of $A(y)$ with eigenvalue $v$,
\begin{align}
&A(y)\,\chi(y;x,z)\,=\,v(y)\,\chi(y;x,z)\,.
 \label{horizontal}
 \end{align}
The function $v(y)$ representing the eigenvalue of $A(y)$ must satisfy $v^2+t(y)=0$,
where
\begin{equation}
t(y)\,=\,-\frac{1}{2}{\rm tr}(A^2(y))\,.
\end{equation}
Using $t(y)$ we define the Seiberg-Witten curve as usual by
\begin{equation}
\Sigma\,=\,\{\,(v,u)\,|\,v^2+t(u)=0\,\}\,.
\end{equation}
 Two linearly independent eigenvectors of $A(y)$ are given by
 \begin{equation}
\chi_\pm(y;x,z)\,=\,\bigg(\begin{matrix}A_0(y)\pm v \\ A_-(y)\end{matrix}\bigg)\,.
 \end{equation}
One of $\chi_{\pm}(y;x,z)$ vanishes at the zeros $u_k$ of $A_-(y)$.
It easily follows from these observations that the coordinates $(x,p)$ and
$(u,v)$ for $\CM_{\rm H}^\ep(C)$ turn into the coordinates for $\CM_{\rm H}(C)$ used in the main text
when $\ep\ra 0$.

It follows from \rf{WKB-horizontal}
that $a_e$ and $k_e$ are in the limit $\ep_2\ra 0$ representable in terms of
periods of
the canonical differential $v$ on $\Sigma$. Given a canonical
basis $\mathbb{B}=\{\al_1,\dots,\al_h;\al_1^{\rm\sst D},\dots,\al_h^{\rm\sst D}\}$
for $H_1'(\Sigma,\BZ)=H_1(\Sigma,\BZ)/H_1(C,\BZ)$  one may define the corresponding periods as
\begin{equation}\label{periods}
a_i=\frac{1}{2\pi}\int_{\al_i}v\,, \qquad
a_i^{\rm\sst D}
=\frac{1}{2\pi}\int_{\al_i^{\rm\sst D}}v\,.
\end{equation}
For given pants decomposition $\si$ one may find a basis $\mathbb{B}_\si$ with the following
property:
For each edge $e$ of $\sigma$ there exists an index $i_e\in\{1,\dots,h\}$ such that the
functions
$a_{i_e}$ and  $a_{i_e}^{\rm\sst D}$ defined in \rf{periods} represent the limits $\ep\ra 0$
of the coordinates
$a_e$ and $k_e$
defined via \rf{FN-FK}, respectively.

The coordinates $a=(a_1,\dots,a_h)$
may be completed into a system of Darboux coordinates $(a,t)$ for $\CM_{\rm\sst H}(C)$
by introducing the coordinates $t=(t_1,\dots,t_h)$ using a variant of the Abel map defined as
\begin{equation}\label{Abel}
t_k\,=\,-\sum_{l=1}^d\int^{u_l}\om_k\,,
\end{equation}
where $\om_k$, $k=1,\dots,h$ are the
Abelian differentials of the first kind on the spectral curve $\Sigma$ which are dual
to the differentials $\al_i$ in the sense that $\int_{\al_i} \om_k=\de_{ik}$.
The functions $t_r$ represent coordinates on the Prym variety.
The fact that the coordinates $(a,t)$  represent Darboux coordinates for $\CM_{\rm\sst H}(C)$
follows from the fact that
\begin{equation}
{\CWL}(a,u,z)\,=\,-\sum_{l=1}^d\int^{u_l}v\,,
\end{equation}
is a generating function for the change of coordinates $(u,v)\leftrightarrow (a,t)$.
Indeed, note that
\begin{equation}
\om_k:=
\frac{1}{2\pi}\,\frac{\pa}{\pa{a_k}}v\,,
\end{equation}
 is an abelian differential on $\Sigma$ satisfying  $\int_{\al_i}\om_k=\de_{ik}$
 as a consequence of \rf{periods}.
We may therefore  conclude that $\CWL(a,u,z)$ satisfies
\begin{equation}
\frac{	1}{2\pi}\,\frac{\pa}{\pa{a_k}}\CWL(a,u,z)=t_k\,,\qquad
\frac{\pa}{\pa u_k}\CWL(a,u,z)=-v_k\,,
\end{equation}
identifying $\CWL(a,u,z)$ as the generating function for the change of coordinates $(u,v)\leftrightarrow (a,t)$.

\section{Classical limits of conformal field theory}\label{Classlim}

We had in the main text
introduced  chiral partition functions $\CZ^{\rm\sst L}_{}(\beta,u,\tau;b)$ and
$\CZ^{\rm\sst WZ}_{}(j,x,\tau;k)$ in Liouville theory and the WZWN model respectively.
It will be helpful to parametrize the representation labels $\beta$ and $j$ appearing in the
arguments of the functions $\CZ^{\rm\sst L}_{}(\beta,u,\tau;b)$ and
$\CZ^{\rm\sst WZ}_{}(j,x,\tau;k)$ as
\begin{align}
&\beta_e\,=\,\frac{Q}{2}+{\mathrm i}\frac{a_e}{\sqrt{\ep_1\ep_2}}\,,\qquad b^2\,=\,\frac{\ep_1}{\ep_2}\,,\\
&j_e\,=\,-\frac{1}{2}+{\mathrm i}\frac{a_e}{\ep_1}\,, \qquad k+2\,=\,-\frac{\ep_2}{\ep_1}\,.
\end{align}
Using this parametrization allows us to introduce chiral partition functions
$\CZ^{\rm\sst L}_{}(a,u,\tau;\ep_1,\ep_2)$
and $\CZ^{\rm\sst WZ}_{}(a,x,\tau;\ep_1,\ep_2)$
depending on two parameters $\ep_1$ and $\ep_2$. We may therefore define two different
classical limits of Liouville theory and the $SL(2)$-WZW model by sending $\ep_1$ or $\ep_2$ to zero,
respectively.
We are interested in the limit where both $\ep_1$ and $\ep_2$ are sent to zero,
but it helps to first study the limit  $\ep_1\ra 0$ with $\ep_2$ finite before sending $\ep_2\ra 0$.
After sending $\ep_1$ to zero we will  find a relation to the moduli space $\CM_{\rm\sst H}^{\ep_2}(C)$
of $\ep_2$-connections.

The two cases related to Virasoro and Kac-Moody algebra, respectively, can be treated in very similar
ways. In each of these cases we will show that the leading asymptotic behavior of the
chiral partition functions,
\begin{equation}\label{ep1to0}
\begin{aligned}
&\log\CZ^{\rm \sst WZ}_{\si}(a,x,\tau;\ep_1,\ep_2)\,\sim\,-\frac{1}{\ep_1}{\CY}^{\rm\sst WZ}_{\si}(a,x,\tau;\ep_2)\,,\\
&\log\CZ^{\rm \sst L}_{\si}(a,u,\tau;\ep_1,\ep_2)\,\sim\,-\frac{1}{\ep_1}{\CY}^{\rm\sst L}_{\si}(a,u,\tau;\ep_2)
\end{aligned}\end{equation}
is represented by functions
$\CY^{\rm\sst WZ}_{}(a,x,\tau;\ep_2)$ and $\CY^{\rm\sst L}_{}(a,u,\tau;\ep_2)$,
which are generating functions for the changes of Darboux variables $(x,p)\leftrightarrow(a,k)$
and $(u,v)\leftrightarrow(a,k)$ for $\CM_{\rm\sst H}^{\ep_2}(C)$, respectively.

The dependence on the variables $x$ (resp. $u$) will be controlled by the partial differential
equations satisfied by  $\CZ^{\rm\sst WZ}_{}(a,x,\tau;\ep_1,\ep_2)$
(resp. $\CZ^{\rm\sst L}_{}(a,u,\tau;\ep_1,\ep_2)$), known as Knizhnik-Zamolodchikov-Bernard (KZB)
and Belavin-Polyakov-Zamolodchikov (BPZ) equations.
In order to control the dependence on the variables $a$
in both cases the crucial tool will be the
Verlinde loop operators defined by integrating the parallel transport defined by KZB- and BPZ-equations,
respectively. The Verlinde loop operators can be represented as difference operators
acting on the $a$-variables. The limit $\ep_1\ra 0$ of the relations between parallel transport and the
corresponding difference operators will govern the $a$-dependence
of $\CY^{\rm\sst WZ}_{}(a,x,\tau;\ep_2)$ and $\CY^{\rm\sst L}_{}(a,u,\tau;\ep_2)$.
The following
discussion considerably refines the previous observations
\cite{Re92,Har} by supplementing the ``other side of the coin''
represented by the Verlinde loop operators.

To simplify the exposition we will spell out the relevant arguments only in the case when $C$ has genus
zero. The dependence on the complex structure of $C$ may then be described using the
positions $z=(z_1,\dots,z_n)$ of the marked points. We will therefore replace the parameters $\tau$
by the variables $z$ in the following.
The generalization of this analysis to higher genus Riemann surfaces will not be too hard.

\subsection{Preparations: Insertions of degenerate fields}

It will be  useful to modify the  conformal blocks by inserting a variable  number of $m$
extra degenerate fields at position $y=(y_1,\dots,y_m)$.

\subsubsection*{WZW model} We will consider conformal blocks of the form
\begin{equation}\label{degins}
\CZ^{\rm\sst WZ}(w,y;x,z)\,:= \big\langle\Phi^{j_n}
(x_n|z_n)\dots\Phi^{j_1}(x_1|z_1)\,
\Phi^{\frac{1}{2}}(w_m|y_m)\dots \Phi^{\frac{1}{2}}(w_1|y_1)
\big\rangle_{C,\vf}.
\end{equation}
We will impose the ``null vector decoupling'' equation on the
degenerate field $\Phi^{\frac{1}{2}}(w|y)$:
\begin{equation}
\pa_w^2\,\Phi^{\frac{1}{2}}(w|y) = 0,
\end{equation}
which means that $\Phi^{+}_{(2,1)}(w|y)$ transforms in the
two-dimensional representation $\C^2 \simeq \C[w]/(w^2)$ of $\fsl_2$. It follows
that $\CZ_{\rm WZ}(w,y;x,z)$ defines
an element $\Psi^{\rm \sst WZ}(y;x,z)$ of $(\BC^2)^{\ot m}$.

The corresponding chiral partition functions $\Psi^{\rm \sst WZ}(y;x,z)$
satisfy additional first order  differential equations governing the $y$-dependence
which will be formulated explicitly below.
The family of chiral partition functions obtained in this way represents
a convenient repackaging of the information contained in
the chiral partition function $\CZ^{\rm \sst WZ}(x,z)$ without extra degenerate fields ($m=0$).
The chiral partition functions $\CZ^{\rm \sst WZ}(x,z)$ essentially represent the boundary
conditions for the integration of the differential equations governing the $y$-dependence of
$\Psi^{\rm \sst WZ}(y;x,z)$. One may recover $\CZ^{\rm \sst WZ}(x,z)$ from the family of
$\Psi^{\rm \sst WZ}(y;x,z)$ by taking suitable limits.
The presence of extra degenerate fields modifies the KZ-equations as
\begin{align}
&-\frac{\ep_2}{\ep_1}\,\frac{\pa}{\pa z_r}\,\Psi^{\rm \sst WZ}(y;x,z)
= \sum_{\substack{r'=1 \\ r'\neq r}}^n\,\eta_{aa'}
\frac{\CJ_r^a\CJ_{r'}^{a'}}{z_r-z_{r'}}\,\Psi^{\rm \sst WZ}(y;x,z)+
\sum_{s=1}^m
\eta_{aa'}\frac{\CJ_r^a{t}^{a'}_s}{z_r-y_s}\,\Psi^{\rm \sst WZ}(y;x,z)\,,
\label{KZheavy}\end{align}
where ${t}^a_s$ denote the matrices representing $\fsl_2$ on the s-th
tensor factor of $(\BC^2)^{\ot m}$, and $\CJ_r^a$ are the
differential operators introduced in \rf{CJdef}.
In addition we get the following $m$ differential equations:
\begin{align}\label{KZlight}
&-\frac{\ep_2}{\ep_1}\,\frac{\pa}{\pa y_s}\,\Psi^{\rm \sst WZ}(y;x,z)
= \sum_{r=1}^n\,\eta_{aa'}\frac{{t}^a_s\CJ_r^{a'}}{y_s-z_r}\,\Psi^{\rm \sst WZ}(y;x,z)+
\sum_{\substack{s=1 \\ s'\neq s}}^n\,\eta_{aa'}
\frac{{t}_s^a{t}_{s'}^{a'}}{z_s-z_{s'}}\,\Psi^{\rm \sst WZ}(y;x,z),\end{align}
The space of solutions to the equations \rf{KZlight} is determined by the space of conformal
blocks without extra degenerate fields $m=0$. This follows from the fact that one may
regard the partition function $\CZ^{\rm \sst WZ}(x,z)$ as initial values for the solution of
\rf{KZlight}. One may, on the other hand, recover the partition functions $\CZ^{\rm \sst WZ}(x,z)$
by considering even $m$ and taking a limit where the insertion points $y_s$ collide
pairwise.

\subsubsection*{Liouville theory}
The situation is similar in the case of Liouville theory. In the
presence of $m$ degenerate fields of weight
$-\frac{1}{2}-\frac{3}{4}b^{-2}$ and $l$ degenerate fields of weight
$-\frac{1}{2}-\frac{3}{4}b^{2}$ the chiral partition functions will
satisfy $l$ BPZ equations \eqref{BPZheavy}
We shall consider the Liouville conformal blocks
\begin{equation}\label{corr++}
\CZ_{\rm\sst L}(y;u,z) \equiv
\left\langle\, \prod_{r=1}^n e^{2\al_r\phi(z_r)}
\prod_{s=1}^{m}e^{-b\phi(y_s)}\prod_{k=1}^l e^{-\frac{1}{b}\phi(u_k)}
\,\right\rangle_{C,\vf}.
\end{equation}
The conformal blocks \rf{corr++} satisfy the null vector
decoupling equations
\begin{subequations}\label{BPZ+}
\begin{align}
& \Bigg( b^2\frac{\pa^2}{\pa u^2_k}+\sum_{r=1}^n
\left(\frac{\De_r}{(u_k-z_r)^2}+\frac{1}{u_k-z_r}\frac{\pa}{\pa z_r}\right)-
\sum_{s=1}^m\left(\frac{3b^2+2}{4(u_k-y_s)^2}-\frac{1}{u_k-y_s}
\frac{\pa}{\pa y_s}\right)\notag\\
& \quad -\sum_{\substack{k'=1\\ k'\neq k}}^l
\left(\frac{3b^{-2}+2}{4(u_k-u_{k'})^2}-\frac{1}{u_k-u_{k'}}
\frac{\pa}{\pa u_{k'}}\right)\Bigg)\CZ_{\rm\sst L}(y;u,z)=0\,,\label{BPZheavy}\\
&\Bigg(
\frac{1}{b^{2}}\frac{\pa^2}{\pa y^2_s}+\sum_{s=1}^n
\left(\frac{\De_r}{(y_s-z_r)^2}+\frac{1}{y_s-z_r}\frac{\pa}{\pa z_r}\right)-
\sum_{k=1}^l \left(\frac{3b^{-2}+2}{4(y_s-u_k)^2}-\frac{1}{y_s-u_k}
\frac{\pa}{\pa u_{k}}\right)\notag\\
& \quad -\sum_{\substack{s'=1\\ s'\neq s}}^m
\left(\frac{3b^2+2}{4(y_s-y_{s'})^2}-\frac{1}{y_s-y_{s'}}
\frac{\pa}{\pa u_{s'}}\right)\Bigg)\CZ_{\rm\sst L}(y;u,z)=0\,.
\label{BPZlight}\end{align}
\end{subequations}
Equations~\rf{BPZ+} imply the fusion rules
\begin{align}
&[V_{-b/2}]\cdot [V_{\alpha}]\,\sim\,
[V_{\alpha-b/2}]+[V_{\alpha-b/2}]\,, \label{furu1}\\
&[V_{-1/2b}]\cdot [V_{\alpha}]\,\sim\,
[V_{\alpha-1/2b}]+[V_{\alpha-/2b}]\,. \label{furu2}
\end{align}

\subsection{Limit $\ep_1\ra 0$}

We will next discuss the behavior of the solutions to the null vector decoupling equations
in the limit $\ep_1\ra 0$.

\subsubsection*{WZW-model}
In order to study the limit $\ep_1\ra 0$ it is useful to
multiply \rf{KZlight} by $\ep_1$ and \rf{KZheavy} by $\ep_1^2$. One may solve the system of
equation \rf{KZheavy}  and \rf{KZlight} with the following ansatz,
\begin{equation}
\Psi^{\rm \sst WZ}(y;x,z)=
e^{-\frac{1}{\ep_1}{\CY}^{\rm\sst WZ}(x,z)}\bigotimes_{s=1}^{n}\psi(y_s;x,z)\big(1+\CO(\ep_1)\big)\,,
\end{equation}
which will yield a solution to  \rf{KZlight} provided $\psi(y;x,z)$ and ${\CY}^{\rm\sst WZ}(x,z)$
satisfy the following system of equations:
\begin{subequations}
\begin{align}\label{KZlight-lim}
&\bigg({\ep_2}\,\frac{\pa}{\pa y}+A(y)\!\bigg)\psi(y;x,z)=0\,,
\end{align}
where
\begin{align}
&A(y)
= \sum_{r=1}^n\,\eta_{aa'}\frac{{t}^a_s\,A_r^{a'}}{y-z_r}\,,\qquad
A_r=\bigg(\begin{matrix}
x_rp_r-l_r & \;2l_rx_r-x_r^2p_r \\ p_r & \;l_r-x_rp_r  \end{matrix}\bigg)\\
& p_r=-\frac{\pa}{\pa x_r}\CY^{\rm\sst WZ}(x,z)\,.
\label{p-from-WZ}\end{align}
We recognize model (B) for the flat connections.
The limit of \rf{KZheavy}  yields in addition
\begin{align}\label{KZheavy-lim}
& H_r:=\ep_2\frac{\pa}{\pa z_r}\CY^{\rm\sst WZ}(x,z)
= \sum_{\substack{r'=1 \\ r'\neq r}}^n\,\eta_{aa'}
\frac{A_r^a \,A_{r'}^{a'}}{z_r-z_{r'}}\,.
\end{align}
\end{subequations}
These equations characterize the Hamiltonians of the Schlesinger
system. We have thereby reproduced results of \cite{Re92,Har}.

\subsubsection*{Liouville theory}

In order to study the limit $\ep_1\ra 0$ it is useful to
multiply \rf{BPZlight} and \rf{BPZheavy} by $\ep_1\ep_2$.
One may solve the system of
equation \rf{BPZheavy}  and \rf{BPZlight} with the following ansatz,
\begin{equation}
\Psi^{\rm \sst L}(y;u,z)=
e^{-\frac{1}{\ep_1}\CY^{\rm \sst L}(u,z)}\prod_{s=1}^{n}\psi^{\rm \sst L}(y;u,z)\big(1+\CO(\ep_1)\big)\,,
\end{equation}
which will yield a solution  \rf{BPZlight} provided
$\psi^{\rm \sst L}(y;u,z)$ and $\CY^{\rm\sst L}(u,z)$
satisfy the following system of equations:
\begin{subequations}\label{BPZlight-lim}
\begin{align}
&
\bigg({\ep_2^2}\frac{\pa^2}{\pa y^2_s}+t(y_s)
\bigg)\psi^{\rm\sst L}(y;u,z)=0\,,
\end{align}
where
\begin{align}
& t(y)=\sum_{s=1}^n
\left(\frac{\de_r}{(y_s-z_r)^2}-\frac{H_r}{y_s-z_r}\right)-\ep_2
\sum_{k=1}^l \left(\frac{3\ep_2}{4(y_s-u_k)^2}-\frac{v_k}{y_s-u_k}\right)\,,\\
&
v_k=-\frac{\pa}{\pa{u_k}}\CY^{\rm\sst L}(u,z)\,, \qquad \de_r=\ep_1\ep_2\De_r\,,
\end{align}\end{subequations}
The equations \rf{BPZheavy} yield in addition
\begin{subequations}\label{BPZheavy-lim}\begin{align}
& v_k^2+t_{k,2}^{}=0\,,\qquad t(y)=\sum_{l=0}^{\infty}t_{k,l}(y-u_k)^{l-2}\,,\\
& H_r=\ep_2\frac{\pa}{\pa{z_r}}\CY^{\rm\sst L}(u,z)\,,\qquad
\end{align}
\end{subequations}
These equations define the Hamiltonians of the Garnier system.

\subsection{Verlinde loop operators}

The dependence of the chiral partition function on the variables $a$ is controlled by the
Verlinde loop operators. They are defined by modifying a conformal block by inserting the vacuum
representation in the form of a pair of degenerate fields, calculating the monodromy of one
of them along a closed curve $\ga$ on $C$, and projecting back to the vacuum representation,
see \cite{AGGTV,DGOT} for more details.
A generating set is identified using pants decompositions.

The calculation of the Verlinde loop operators is almost a straightforward extension of what has been
done in the literature. The necessary results have been obtained in \cite{AGGTV,DGOT} for Liouville
theory without extra insertions of degenerate fields $V_{-b/2}(y)$. It would be straightforward to
generalize these observations to the cases of our interest. For the case of Kac-Moody conformal
blocks one could assemble the results from the known fusion and braiding matrices of
an extra degenerate field $\Phi^{\frac{1}{2}}(w,y)$. As a shortcut let us note, however, that
the results relevant for the problem of our interest, the limit $\ep_1\ra 0$, can be obtained in a
simpler way.

One may start on the Liouville side.
The key observation to be made is the fact that the presence of
extra degenerate fields  $V_{-1/2b}(y)$ modifies  the monodromies of $V_{-b/2}(y)$
only by overall signs, as the monodromy of $V_{-b/2}(y)$ around
$V_{-1/2b}(u_k)$ is equal to minus the identity. It is useful to observe (see Appendix \ref{DEG-SOV})
that the separation of variables transformation
maps the degenerate field $\Phi^{\frac{1}{2}}(w,y)$
to the degenerate field $V_{-b/2}(y)$. It follows that
the monodromies of $\Phi^{\frac{1}{2}}(w,y)$ must coincide with the monodromies of
$V_{-b/2}(y)$ up to signs.
Using the results of \cite{AGGTV,DGOT} we conclude that
\begin{equation}\label{Verlindeact}\begin{aligned}
&(\pi^{\rm\sst V}(\ga_{e,s})\CZ^{\rm\sst WZ})(a,u,z)\,=\, \nu_{e,s}\,\SL_{e,s}\cdot\CZ^{\rm\sst WZ}(a,u,z)\,,\\
&(\pi^{\rm\sst V}(\ga_{e,t})\CZ^{\rm\sst WZ})(a,u,z)\,=\, \nu_{e,t}\,\SL_{e,t}\cdot\CZ^{\rm\sst WZ}(a,u,z)\,,
\end{aligned}
\end{equation}
where $\nu_{e,s}\in\{\pm 1\}$ and $\nu_{e,r}\in\{\pm 1\}$,
while the explicit expressions for the difference operators $\SL_{e,s}$, $\SL_{e,t}$ are
\begin{subequations}\label{Verl-op}
\begin{align}
   \SL_{e,s}&=2\cosh(2\pi \sa_e/\ep_2)\,. \\
    \SL_{e,t}&=   
\frac{2\cos(\pi \ep_1/\ep_2)(\SL_{e,2}\SL_{e,3}+\SL_{e,1}\SL_{e,4})+
\SL_{e,s} (\SL_{e,1}\SL_{e,3}+\SL_{e,2}\SL_{e,4})}
{2\sinh \big(\frac{2\pi}{\ep_2}(\sa_e+\frac{\mathrm{i}}{2}\ep_1)\big)
 2\sinh \big(\frac{2\pi}{\ep_2}(\sa_e-\frac{\mathrm{i}}{2}\ep_1)\big)}
\\ \label{quantum't Hooft}
& \quad +  \sum_{\xi=\pm 1}
\frac{1}{\sqrt{2\sinh(2\pi\sa_e/\ep_2)}}
e^{{\pi}\xi\sk_e/\ep_2}
\frac{\sqrt{c_{12}(\SL_{r,s})c_{34}(\SL_{r,s})}}{2\sinh(2\pi\sa_e/\ep_2)}
e^{{\pi}\xi\sk_e/\ep_2}
\frac{1}{\sqrt{2\sinh(2\pi\sa_e/\ep_2)}}\,, \nonumber
\end{align}
\end{subequations}
using the notation $c_{ij}(\SL_{e,s})=\SL_{e,s}^2+\SL_{e,i}^2+\SL_{e,j}^2+\SL_{e,s}^{} \SL_{e,i}^{}\SL_{e,j}^{}-4$, and
\begin{equation}
\sk_e\,=\,\frac{\ep_1\ep_2}{2\pi \mathrm{i}}\frac{\pa}{\pa a_e}\,.
\end{equation}

As the KZB-equations \rf{KZlight} turn into the horizontality condition \rf{KZlight-lim}, the Verlinde
loop operators will turn into trace functions when $\ep_1\ra 0$. The limit of the left hand side of \rf{Verlindeact}
is therefore  found by replacing  $\pi^{\rm\sst V}(\ga_{e,s})$ and $\pi^{\rm\sst V}(\ga_{e,t})$ with the
expressions in \rf{FN-FK}, calculated from the connection $A(y)$ appearing in \rf{KZlight-lim}.
Note that the connection $A(y)$ is thereby defined as a function of the parameters $x$ and $a$.
The limit $\ep_1\ra 0$ of the right hand side of \rf{Verlindeact} is straightforward to analyze by using
\rf{ep1to0} and \rf{Verl-op}. It can be  expressed
in terms of the derivative of $\CY^{\rm\sst WZ}$ with respect to the
variable $a$. In this way one finds that the
the limit $\ep_1\ra
0$ of equations \rf{Verlindeact} implies the relations
\begin{equation}\label{Verl-lim}
k_e(a,u)\,=\,\ep_2\frac{\mathrm{i}}{2\pi }\frac{\pa}{\pa a_e} \CY^{\rm\sst WZ}(a,u,z)\,.
\end{equation}
Equation \rf{Verl-lim} identifies $\CY^{\rm\sst WZ}(a,u,z)$ as the generating function for the change of
variables $(x,p)\leftrightarrow (a,k)$. The analysis in the Liouville case is very similar.

\subsection{Limit $\ep_2\ra 0$}

It remains to discuss the behavior in the limit $\ep_2\ra 0$
of $\CY^{\rm\sst WZ}_{}(a,x,z;\ep_2)$ and $\CY^{\rm\sst L}_{}(a,u,z;\ep_2)$.
We claim that in the two cases we find a behavior of the form
\begin{align}\label{CYlim}
& {\CY}^{\rm\sst WZ}(a,x,z)\,\sim\,\frac{1}{\ep_2}\CF^{\rm\sst
  WZ}(a,z)+\tilde \CW^{\rm\sst WZ}(a,x,z)+\dots\,,\\
& {\CY}^{\rm\sst L}(a,u,z)\,\sim\,\frac{1}{\ep_2}\CF^{\rm\sst
  L}(a,z)+\tilde \CW^{\rm\sst L}(a,u,z)+\dots\,,
\label{ep2-lim-L}\end{align}
where $\CF^{\rm\sst WZ}(a,z)= \CF^{\rm\sst L}(a,z)$, while
$\tilde \CW^{\rm\sst WZ}(a,x,z)$ and
$\tilde \CW^{\rm\sst L}(a,x,z)$ are the generating functions for the changes of
variables $(x,p)\leftrightarrow (a,t)$ and $(u,v)\leftrightarrow (a,t)$, respectively.


We begin by considering \rf{ep2-lim-L}.
The equation
\rf{BPZlight-lim} can be solved to leading order  by a WKB-ansatz
\begin{equation}\label{BPZlight-WKB}
\psi^{\rm\sst L}(y;u,z)\,\asymp\,e^{-\frac{1}{\ep_2}\int^{y}du\,v(u)}\,,\qquad
(v(y))^2+t(y)=0\,.
\end{equation}

The asymptotics of the
generating function $\CY^{\rm\sst L}(a,u;z)$
which coincides with the classical Liouville conformal blocks
will be of the form
\begin{align}\label{CYlim1}
& {\CY}^{\rm\sst L}(a,u,z)\,\sim\,\frac{1}{\ep_2}\CF^{\rm\sst
  L}(a,z)+\tilde \CW^{\rm\sst L}(a,u,z)+\dots\,,\,.
\end{align}
Indeed, an expansion of the form will satisfy \rf{BPZheavy-lim} and \rf{Verl-lim} if
$\CF_{\rm\sst L}(a,z)$ satisfies
\begin{equation}
\frac{\pa}{\pa z_r}\CF^{\rm\sst L}(a,z)\,=\,H_r\,,\qquad
\frac{\mathrm{i}}{2\pi }\frac{\pa}{\pa a_e}\CF^{\rm\sst L}(a,z)\,=\,a_e^{\rm\sst D}\,,
\end{equation}
identifying $\CF(a,z)$ as the prepotential, and if furthermore
\begin{equation}\label{W'u}
\frac{\pa}{\pa u_k} \tilde \CW^{\rm\sst L}(a,u,z)\,=\,-v_k\,.
\end{equation}
This means that
\[
\CWL(a,u;z)\,=\,-\sum_{l=1}^d\int^{u_l}v\,.
\]
Following the discussion in Appendix \ref{Conn-to-Higgs} we
may identify $\tilde \CW^{\rm\sst L}(a,u;z)$ as the
generating function of the standard change of
Darboux variables $(u,v)\leftrightarrow (a,t)$
which is defined by the Abel map.

The corresponding statement for $\tilde \CW^{\rm\sst WZ}(a,x;z)$ now follows easily from
\rf{p-from-WZ}, and the fact that ${\CY}^{\rm\sst WZ}(a,x,z)$ and
${\CY}^{\rm\sst L}(a,u,z)$ differ only by the generating function
${\CY}^{\rm\sst SOV}(x;u,z)$ for the change of Darboux variables $(x,p)\leftrightarrow (u,v)$
which  does not depend on $a$.

\section{Explicit relation between Kac-Moody and Virasoro conformal
  blocks}\label{App-expl}

We will explain in this appendix how to obtain an explicit integral
transformation between the conformal blocks in Liouville theory and in
the WZW model using the observations made in Section
\ref{Liou-WZW}. This is the separation of variables (SOV) relation
\eqref{SOV-1} which we discussed in the Introduction.

\subsection{SOV transformation for conformal blocks}

In order to partially fix the global $\fsl_2$-constraints we
shall send $z_n\ra \infty$ and $x_n\ra\infty$, defining the reduced
conformal blocks $\check{\CZ}^{\rm\sst WZ}(x,z)$ which depend on
$x=(x_1,\dots,x_{n-1})$ and $z=(z_1,\dots,z_{n-1})$.
Let $\tilde{\CZ}^{\rm\sst WZ}(\mu,z)$ be the Fourier-transformation
of the reduced
conformal block $\check{\CZ}^{\rm\sst WZ}(x,z)$ of the WZW model w.r.t. the
variables $x$. It depends on $\mu=(\mu_1,\dots,\mu_{n-1})$ subject
to $\sum_{r=1}^{n-1}\mu_r=0$. There then exists a solution $\CZ^{\rm\sst L}(y,z)$  to the BPZ-equations
\begin{equation}\label{BPZ}
\mathcal{D}_{u_k}^{{\rm\sst BPZ}}\cdot \CZ^{\rm\sst L} = 0,\quad \forall k=1,\ldots,l,
\end{equation}
with differential operators $\mathcal{D}_{u_k}^{{\rm\sst BPZ}}$ given as
\begin{align*}
\mathcal{D}_{u_k}^{{\rm\sst BPZ}} &= b^{2}\frac{\pa^2}{\pa u_k^2}
+\sum_{r=1}^n \left(\frac{\De_r}{(u_k-z_r)^2} + \frac{1}{u_k-z_r}\frac{\pa}{\pa z_r}\right)
-\sum_{\substack{k'=1\\ k'\neq k}}^l
\left(\frac{3b^{-2}+2}{4(u_k-u_{k'})^2}-\frac{1}{u_k-u_{k'}}
\frac{\pa}{\pa u_{k'}}\right),
\end{align*}
such that the following relation holds
\begin{equation}\label{main}
\tilde{\CZ}^{\rm\sst WZ}(\mu,z) =  u_0\,\delta\big({\textstyle  \sum_{i=1}^{n-1}\mu_i}\big)\,
\Theta_n(y,z)\, \CZ^{\rm\sst L}(y,z)\,.
\end{equation}
 The function $\Theta_n(y,z)$ that appears in this relation is defined
as
\begin{equation}
\Theta_n(y,z) =
\prod_{r<s\leq n-1} (z_{r}-z_s)^{\frac{1}{2b^2}}
\prod_{k<l\leq n-3} (u_{k}-u_l)^{\frac{1}{2b^2}}
\prod_{r=1}^{n-1}\prod_{k=1}^{n-3}(z_r-u_k)^{-\frac{1}{2b^2}}.
\label{thetan}
\end{equation}
The relation \rf{main}
will hold provided that the respective variables
are related as follows:
\begin{enumerate}
\item[(1)] The variables $\mu_1,\ldots,\mu_{n-1}$ are related
to $u_1,\ldots,u_{n-3},u_0$ via
\begin{equation}\label{magic}
\sum_{r=1}^{n-1} \frac{\mu_r}{t-z_r} =
u_0\frac{\prod_{k=1}^{n-3}(t-u_k)}{\prod_{r=1}^{n-1} (t-z_r)} .
\end{equation}
In particular, since $\sum_{r=1}^{n-1} \mu_r=0$, we
have $u_0=\sum_{r=1}^{n-1} \mu_r z_r$.
\item[(2)] $b^2=-(k+2)^{-1}$.
\item[(3)] The
Liouville momenta are given by
\begin{equation}\label{alpha-j}
\a_r\equiv\a(j_r):=b(j_r+1)+ \frac{1}{2b}.
\end{equation}
\end{enumerate}
We may use formula \rf{main} to construct bases of solution to the
KZ-equations from Liouville conformal blocks.

\subsection{Reformulation as integral transformation}

We want to write the expression for $\check{\CZ}^{\rm\sst WZ}(x,z)$
\begin{align}
\check{\CZ}^{\rm\sst WZ}(x,z)\,=\,\int
\frac{d\mu_1}{\mu_1}\dots \frac{d\mu_{n-1}}{\mu_{n-1}}\,
\de\big({\textstyle \sum_{r=1}^{n-1}\mu_r}\big)\,\Theta_n(u,x)\,\CZ_{\rm\sst L}(u,z)
\prod_{r=1}^{n-1}\mu_r^{-j_r}e^{i\mu_r x_r}\,,
\label{SOVxx}
\end{align}
as explicitly as possible. To this aim let us note first that
\begin{align}
\mu_r(u)\,=\,u_0\la_r(u)\,,
\qquad\la_r(u):=\frac{\prod_{k=1}^{n-3}(z_r-u_k)}{\prod_{s\neq r}^{n-1}
(z_r-z_s)}\,,
\end{align}
and furthermore
\begin{align}
& \frac{d\mu_1}{\mu_1}\dots \frac{d\mu_{n-1}}{\mu_{n-1}}\,
\de\big({\textstyle \sum_{r=1}^{n-1}\mu_r}\big)\,\Theta_n(u|x)\,
=\,\frac{du_0}{u_0}\,d\nu(u)\,,\\
& d\nu(u)\,:=\,du_1\dots du_{n-3}
\prod_{r\neq s}^{n-1}(z_r-z_s)^{1+\frac{1}{2b^2}}
\prod_{r=1}^{n-1}\prod_{k=1}^{n-3}(z_r-u_k)^{-1-\frac{1}{2b^2}}
\prod_{k<l}^{n-3}(u_k-u_l)^{1+\frac{1}{2b^2}}\notag \\
& \qquad\quad = du_1\dots du_{n-3}\;
\prod_{r=1}^{n-1}[\la_r(u)]^{-1-\frac{1}{2b^2}}
\prod_{k<l}^{n-3}(u_k-u_l)^{1+\frac{1}{2b^2}}\,.
\notag\end{align}
We may therefore calculate
\begin{align}
\check{\CZ}^{\rm\sst WZ}(x,z)\,=\,
\int d\nu(u) \;\prod_{r=1}^{n-1}\la_r^{-j_r}\int \frac{du_0}{u_0}\;
u_0^{-J}\,\CZ_{\rm\sst L}(u,z)\prod_{r=1}^{n-1}e^{iu_0 \la_r x_r}
\,,
\label{SOVxxx}
\end{align}
where $J:=-j_n+\sum_{r=1}^{n-1}j_r$.
The integral over $u_0$ is of the form
\begin{equation}
\int \frac{du_0}{u_0}\;
u_0^{-J} \prod_{r=1}^{n-1}e^{iu_0\la_r x_r}
\,=\,N_J\,\Bigg( \sum_{r=1}^{n-1}\la_r x_r\Bigg)^J\,,
\end{equation}
where $N_J$ depends neither on $x$ nor on $z$. It follows that
\begin{align}
\check{\CZ}^{\rm\sst WZ}(x,z) &\,=\,N_J\int d\nu(u) \;\left({\textstyle \sum_{r=1}^{n-1}\la_r x_r}\right)^J
\CZ_{\rm\sst L}(u,z)\prod_{r=1}^{n-1}\la_r^{-j_r}\,,\notag \\
&\,=\,N_J\int du_1\dots du_{n-3}\;\CK^{\rm\sst SOV}(x,u)\,
\CZ^{\rm\sst L}(u,z)\,,
\label{SOVtrsf}\end{align}
where the kernel $\CK^{\rm\sst SOV}(x,u)$ is defined as
\begin{align}\label{KSOVdef}
&\CK^{\rm\sst SOV}(x,u):=\\
&\qquad=\left[\,{\sum_{r=1}^{n-1}x_r
\frac{\prod_{k=1}^{n-3}(z_r-u_k)}{\prod_{s\neq r}^{n-1}
(z_r-z_s)}}\right]^J\,
\prod_{k<l}^{n-3}(u_k-u_l)^{1+\frac{1}{2b^2}}
\prod_{r=1}^{n-1}\Bigg[\frac{\prod_{s\neq r}^{n-1}
(z_r-z_s)}{\prod_{k=1}^{n-3}(z_r-u_k)}\Bigg]^{\al_r/b}\,.
\notag\end{align}
Note that the $x$-dependence it entirely in the first factor on
the right hand side of \rf{KSOVdef}.

The choice of contours in \rf{SOVtrsf} is a delicate issue that we will not address here.
Using the standard contour $\mathbb R$ in the definition of the Fourier-transformations in \rf{SOVxx}
will of course determine a particular choice of contours in  \rf{SOVtrsf}. Any choice of contours
that ensures absence of boundary terms in the relation between the differential equations satisfied by
$\check{\CZ}^{\rm\sst WZ}(x,z)$ and $\CZ^{\rm\sst L}(u,z)$ could also be
taken to define a relation
of the form \rf{SOVtrsf} between bases of conformal blocks in the WZW-model and in Liouville theory.
Changing the contours in \rf{SOVtrsf}
amounts to a change of basis in the space of solutions to the KZ-equations obtained from a fixed basis
in the space of Liouville conformal blocks. It would be interesting to identify the basis defined
by \rf{SOVtrsf} for a given choice of contours  precisely, and to investigate the dependence on the
choice of contours.

\subsection{Semiclassical limit}

Now we consider semiclassical limit $\ep_1,\ep_2\ra 0$, setting
\begin{equation}
\al_r\,=\,(\ep_1\ep_2)^{-\frac{1}{2}}\,l_r\,.
\end{equation}
We have then
\begin{equation}\label{epto0''}
  \log\CK^{\rm\sst SOV}(x,u)\,=\,\ep_1^{-1}
   \tilde \CW^{\rm\sst SOV}(x,u)+\CO(\ep_1^0)\,,
\end{equation}
with
\begin{align}
\tilde \CW^{\rm\sst SOV}(x,u)\,=\,&\,\kappa\log\left[\,\sum_{r=1}^{n-1}x_r
\frac{\prod_{k=1}^{n-3}(z_r-u_k)}{\prod_{s\neq r}^{n-1} (z_r-z_s)}
\right]\\
& \qquad+\sum_{r=1}^{n-1}l_r
\Bigg[\sum_{s\neq r}^{n-1}
\log(z_r-z_s)-\sum_{k=1}^{n-3}\log(z_r-u_k)\Bigg]\,.
\notag\end{align}
We have denoted $\kappa:=-l_n+\sum_{r=1}^{n-1}l_r$.
If we send only $\ep_1\ra 0$, we get a modified result:
\begin{equation}
\log\CK^{\rm\sst SOV}(x,u)\,=\,\ep_1^{-1} \tilde \CW^{\rm\sst SOV}(x,u;\ep_2)+\CO(\ep_1^0)\,
\end{equation}
with
\begin{align}
\tilde \CW^{\rm\sst SOV}(x,u;\ep_2)\,=\,&\,\kappa\log\left[\,\sum_{r=1}^{n-1}x_r
\frac{\prod_{k=1}^{n-3}(z_r-u_k)}{\prod_{s\neq r}^{n-1} (z_r-z_s)}
\right]+\frac{\ep_2}{2}\sum_{k<l}^{n-3}\log(u_k-u_l)\\
& \qquad+\sum_{r=1}^{n-1}l_r
\Bigg[\sum_{s\neq r}^{n-1}
\log(z_r-z_s)-\sum_{k=1}^{n-3}\log(z_r-u_k)\Bigg]\,.
\notag\end{align}

\subsection{SOV transformation in the presence of degenerate
  fields}\label{DEG-SOV}

We also use the  version of this correspondence in the presence of the
fields $\Phi^{\frac{1}{2}}(w|y)$, as appear in \rf{degins}.
This is kind of interesting. Note that the Fourier-transformation of the
null vector equations $\pa_{w}^2\Phi^{\frac{1}{2}}(w|y)=0$
gives $\mu^2\tilde\Phi^{\frac{1}{2}}(\mu|y)=0$. This indicates that
conformal blocks containing
$\tilde\Phi^{\frac{1}{2}}(\mu|y)$ must be understood as distributions
with support at $\mu=0$. If we send $\mu_r\ra 0$ in the change of
variables \rf{magic}, we will loose the pole at $t=z_r$
on the left hand side.
This means that one $u_k$ must approach $z_r$ in order to cancel the
pole at $t=z_r$
on the right hand side of \rf{magic}. It follows that the degenerate
field $e^{-b^{-1}\phi(u_k)}$ fuses with the field $e^{2\al_r\phi(z_r)}$.
Applying these observations to the case where $j_r=1/2$, which corresponds
to $\al_r=\frac{1}{2}Q+b$ we get as leading term in the
OPE of $e^{-b^{-1}\phi(u_k)}e^{2\al_r\phi(z_r)}$ a field
with conformal dimension $\De_{-b/2}$, which is degenerate.
This indicates that the WZW conformal blocks \rf{degins}
can be represented in terms of the Liouville conformal
blocks \rf{corr++} with $l=n-3$, where the insertion of a
field $\Phi^{\frac{1}{2}}(w_s|y_s)$
corresponds to the insertion of
$e^{-b\phi(y_s)}$. Even if the argument above may look
delicate, the conclusion seems hard to avoid: We need to map
the field $\Phi^{\frac{1}{2}}(w|y)$ to another field with two-dimensional monodromy.
The only candidate with the right behavior for $b\ra 0$ is
$e^{-b\phi(y_s)}$.


\bibliographystyle{JHEP_TD}
\bibliography{SOV-04}

\end{document}